\documentclass[12pt]{article}
\usepackage{amscd}
\usepackage{verbatim}
\usepackage{hyperref}
\usepackage{amssymb}
\usepackage{amsmath}
\usepackage{tabularx}
\begin{document}
\begin{center}
\vspace{24pt} { \large \bf A note on the action with the Schwarzian at the stretched horizon} \\
\vspace{30pt}
\vspace{30pt}
\vspace{30pt}
{\bf Mohd Ali \footnote{mohd.ali@students.iiserpune.ac.in}}, {\bf Vardarajan
Suneeta\footnote{suneeta@iiserpune.ac.in}}\\
\vspace{24pt} 
{\em  The Indian Institute of Science Education and Research (IISER),\\
Pune, India - 411008.}
\end{center}
\date{\today}
\bigskip
\begin{center}
{\bf Abstract}
\end{center}
In this paper, we discuss the quantization of an interesting model of Carlip which appeared recently. It shows a way to associate boundary degrees of freedom to the stretched horizon of a stationary non-extremal black hole, as has been done in JT gravity for near-extremal black holes. The path integral now contains an integral over the boundary degrees of freedom, which are time reparametrizations of the stretched horizon keeping its length fixed. These boundary degrees of freedom can be viewed as elements of $Diff(S^1)/S^1$, which is the coadjoint orbit of an ordinary coadjoint vector under the action of the Virasoro group. From the symplectic form on this manifold, we obtain the measure in the boundary path integral. Doing a one-loop computation about the classical solution, we find that the one-loop answer is not finite, signalling that either the classical solution is unstable or there is an indefiniteness problem with this action, similar to the conformal mode problem in quantum gravity. Upon analytically continuing the field, the boundary partition function we get is independent of the inverse temperature and does not contribute to the thermodynamics at least at one-loop. This is in contrast to the study of near-extremal black holes in JT gravity, where the entire contribution to thermodynamics is from boundary degrees of freedom.

\newpage

\section{Introduction}

In recent years, the connection between dimensionally reduced four dimensional near-extremal black holes and Jackiw-Teitelboim (JT) gravity \cite{JT},\cite{JT1} in two dimensions has been extensively explored in \cite{maldacena}, \cite{almheiri},\cite{trivedi}, \cite{trivedi1}, \cite{trivedi2}, \cite{iliesiu2}, \cite{iliesiu3} \footnote{Connections between JT gravity and other nonsupersymmetric black holes have been discussed in \cite{larsen}, \cite{larsen1}. Connections between near-extremal black holes and dimensionally reduced higher derivative gravity are discussed in \cite{nabamita}. } . In particular, in many of these examples, the path integral for the dimensionally reduced system becomes that of a boundary theory with action which is a Schwarzian derivative of time reparametrizations (a review of these developments can be found in \cite{sarosi}). These boundary degrees of freedom contribute non-trivially to the thermodynamics. In a very interesting development, Carlip \cite{carlip} has argued that under certain choice of boundary conditions, the dimensionally reduced near-horizon region of an arbitrary stationary non-extremal black hole also gives rise to a Schwarzian-like action on the stretched horizon, treating the stretched horizon as a boundary. The action differs from the usual Schwarzian action by a pre-factor. Nevertheless, it leads to the question of whether we can generalize results in JT gravity for near-extremal black holes to stationary non-extremal black holes. In particular, are there boundary degrees of freedom that can be associated with the stretched horizon for a stationary non-extremal black hole similar to near-horizon degrees of freedom for near-extremal black holes? Carlip's proposal can also be seen as a first step in this direction. Other approaches to flat space JT gravity can be found in \cite{oblak}.

In this paper, we analyze and attempt to quantize two boundary actions corresponding to two different boundary conditions in Carlip's paper. More precisely, we are interested in obtaining one-loop corrections to the classical actions. In the context of JT gravity and near-extremal black holes, one-loop corrections to the Schwarzian action have been obtained and the Schwarzian action was shown to be one-loop exact \cite{stanfordwitten}. This used the fact that the time reparametrization is an element of $Diff(S^1)/SL(2,R)$ which is the orbit of a particular coadjoint vector under the action of the Virasoro group (the exceptional orbit). Thus $Diff(S^1)/SL(2,R)$ is a symplectic manifold and there is a natural symplectic measure that can be used in the one-loop computation. Further, results from mathematics (the Duistermaat-Heckman formula) imply that the Schwarzian theory is one-loop exact. We argue that in Carlip's model, the boundary degrees of freedom can be seen as elements of $Diff(S^1)/S^1$ which is the orbit of an ordinary coadjoint vector under the action of the Virasoro group (the ordinary orbit). This is also a symplectic manifold and we can use the symplectic measure on this manifold to compute one-loop corrections to the classical solutions to Carlip's actions. The measure depends on two constants, the coadjoint vector and the central charge. When we attempt to compute the one-loop correction about the classical solution, we find the correction is divergent due to Gaussian-like integrals with the wrong sign. This is indicative of either the classical solution being unstable, or a problem due to the indefiniteness of this particular boundary action similar to the conformal mode problem in Euclidean quantum gravity. Since the action has higher derivative terms, we are unable to find more than one classical solution easily and therefore, it is not possible to determine if this problem occurs for other classical solutions. However, we can certainly do an analytic continuation of the perturbation to imaginary values, similar to that done in Euclidean quantum gravity by Gibbons, Hawking and Perry \cite{GHP}. Upon doing this, we find that the correction to the partition function is independent of $\beta$, the inverse temperature and thus, this correction does not contribute to thermodynamics. This is different from that obtained in JT gravity for the quantization of the Schwarzian action. There, the boundary contribution was the only contribution and it had a non-trivial dependence on the inverse temperature $\beta$. Thus we come to the conclusion that either the classical solution is unstable or, doing the analytic continuation, the one loop correction does not affect thermodynamics. There does not seem to be any way for horizon degrees of freedom to contribute to the thermodynamics of the non-extremal black hole in the way they do for near-extremal black holes using the Schwarzian theory.

The outline of the paper is as follows: In section II, we summarize some of the key assumptions and results of Carlip's model to set the stage for computations in following sections. In section III, we motivate and describe how to quantize the boundary actions of Carlip from the previous section. In section IV, we describe the reparametrizations of the stretched horizon as elements of the ordinary orbit of a set of coadjoint vectors under the action of the Virasoro group. This manifold is $Diff(S^1)/S^1$, which is symplectic. We write the symplectic form on this space whose Pfaffian gives the symplectic measure on this manifold. We use this measure while computing the one-loop correction. In section V, we discuss the classical solutions to Carlip's actions and compute the classical contribution of the actions. In sections VI and VII, we consider perturbations about the classical solutions and the one-loop corrections. In section VIII, we discuss possible generalizations of this model and quantization. In section IX, we discuss the dilaton as an external coupling in the boundary action. Section X provides a summary and discussion of the results in this paper.
In Appendix A, we discuss classical solutions to dilaton gravity. In Appendix B, we discuss the corrections to Carlip's actions coming from the induced metric computed using the exact two-dimensional classical metric in spherically symmetric reduced gravity in comparison to the flat metric as Carlip has done. The corrections are higher order in the length of the stretched horizon (the small parameter in our problem), than what we have considered in this paper. Appendix C deals with some technical details of the symplectic form on the symplectic manifold. In Appendix D, we write the one-loop corrections around an arbitrary classical solution. Finally, in Appendix E, we write results in zeta function regularization needed to compute the one-loop corrections to the classical solution.

\section{Summary of results in Carlip's work}
We review certain assumptions and results in Carlip's paper \cite{carlip} to set the stage for the computations in subsequent sections.
Consider the general metric of a $D+1$ dimensional stationary black hole written in the form
\begin{align}
ds^2 = \tilde g_{ab}dx^a dx^b + \phi_{\mu\nu}(dy^\mu + A_a{}^\mu dx^a)(dy^\nu + A_b{}^\nu dx^b) \, ,
\label{a1}
\end{align}
Here, $a,b = 0,1$ label the $r-t$ plane while Greek indices run from 2,...,D-1 and label the coordinates transverse to the r-t plane.
As shown by Carlip in \cite{carlip2}, the dimensionally reduced near-horizon action is
\begin{align}
S = \frac{1}{16\pi G}\int_M\!d^2x\,\sqrt{g}\left\{ \varphi R  + V[\varphi]\right\} + \dots \, ,
\label{a2}
\end{align}
where
\begin{equation}
\varphi = \sqrt{|\det\phi_{\mu\nu}|}.
\label{a2a}
\end{equation}
\begin{equation}
\tilde g_{ab} = \varphi^{\frac{-(D-3)}{D-2}} g_{ab},
\label{a2b}
\end{equation}
 and the scalar curvature in (\ref{a2}) is with respect to $g_{ab}$.
The ellipses denote the fact that the near horizon action has additional terms with Kaluza Klein fields from the dimensional reduction. The action also has a (dimensionally reduced) Gibbons-Hawking-York (GHY) boundary term when the metric is fixed at the boundary, given by
\begin{align}
S_{GHY} = \frac{1}{8\pi G}\int_{\partial M}\!dx\sqrt{h}\,\varphi K   \, ,
\label{a3}
\end{align}
The equations of motion arising from the near-horizon action (\ref{a2}) computed in \cite{carlip2} are (ignoring matter-like fields which arise in dimensional reduction):
\begin{eqnarray}
&&\nabla_a \nabla_b \varphi - g_{ab} \nabla^c \nabla_c \varphi + (1/2)~ g_{ab} V = 8\pi G T_{ab}; \nonumber \\
&& R + dV/d\varphi = 0.
\label{a3a}
\end{eqnarray}
Here, $T_{ab}$ is the matter stress-energy tensor corresponding to some other fields which are decoupled from $\varphi$. In what follows, we will not consider these other fields for simplicity, and our $T_{ab} = 0$. We will also not consider other fields arising from dimensional reduction in the analysis in this paper. The contribution of some of these matter-like fields in the action is zero in the case of dimensional reduction of the Schwarzschild or Reissner-Nordstrom black holes. Further, when we go sufficiently close to the horizon, the terms in the action involving the fields arising from dimensional reduction simplifies considerably. For dimensionally reduced spherically symmetric gravity $V(\varphi)=\frac{1}{2\sqrt{\varphi}}$, we have the general solution \cite{Cavaglia}
\begin{equation}\label{a3a1}
ds_{E}^2=\frac{1}{2} \Big( (r-r_+)d\tau^2+\frac{r^2}{r-r_+}dr^2 \Big)
\end{equation}
where $\varphi=r^2/4$ and $r_+=2GM$ is the black hole horizon. Near the horizon, this metric simplifies approximately to the flat metric.

In fact, \emph{sufficiently} close to the horizon of any stationary, non-extremal black hole, the two dimensional metric $\tilde g_{ab}$ is flat (Rindler) \cite{carlip2}. Consider the Schwarzschild metric. The $r-t$ plane part of the metric can be written sufficiently close to the horizon as the flat metric in the form (\ref{a5a}) below. To see this, we Wick rotate the Schwarzschild time coordinate $\tau = it$, and change from the Schwarzschild radial coordinate $r$ to $\rho$ as
\begin{align}
 r = r_+ + \frac{1}{2}\kappa\rho^2 ,
 \label{a5}
 \end{align}
 where $\kappa$ is the surface gravity.
Here, $r_+$ is the radius of the horizon. $\tau$, the Euclidean time has been defined such that it is periodic with period $\beta = \frac{2\pi}{\kappa}$. So $\tau \sim \tau + \beta$. We note that this differs from the $\tau$ in Carlip's paper by a factor. In his paper, the coordinate has period $2\pi$.
We then end up with the flat metric
\begin{align}
ds^2 = d\rho^2 + \kappa^2 \rho^2d\tau^2 \, .
\label{a5a}
\end{align}
This is not in contradiction with the equations (\ref{a3a}) because sufficiently close to the horizon, the dilaton is a constant and the potential $V(\varphi) \sim k$ where $k$ is a constant. Then the flat metric is indeed a solution to (\ref{a3a}).
Of course, the metric receives corrections which are of $O(\rho^4)$, but to leading order in $\rho$ we can assume the metric to be (\ref{a5a}). Also, in the Schwarzschild context, sufficiently close to the horizon, this metric is $\tilde g_{ab}$ which differs from $g_{ab}$ by a conformal factor. The conformal factor is trivial at leading order in $\rho$ due to the boundary conditions we place on the dilaton, as we see below.
In these coordinates, the horizon is at $\rho = 0$. We now consider a stretched horizon in the $\rho-\tau$ plane, given by the curve $(\rho(\sigma), \tau(\sigma))$, where $\sigma$ is a parameter with range $[0,2\pi]$. We denote the stretched horizon by $\Delta$. Its length is
\begin{align}
\ell = \oint\!ds = \int_{0}^{2\pi}\!\varepsilon \,d\sigma \quad \hbox{with}\quad
\varepsilon = \frac{ds}{d\sigma} = \left(\rho^{\prime 2} + \kappa^2 \rho^2\tau^{\prime 2}\right)^{1/2} \, ,
\label{a6}
\end{align}
Carlip chooses a parameter $\sigma$ such that $\varepsilon$ is constant on the stretched horizon. This closely parallels how the boundary curve in the near-horizon region is parametrized in JT gravity \cite{maldacena}. $\varepsilon$ is taken small as the stretched horizon is close to the actual horizon at $\rho = 0$. Crucially, Carlip also assumes $\rho$ is $\mathcal{O}(\varepsilon)$ while
$\rho'$ is $\mathcal{O}(\varepsilon^2)$.

If $\varphi$ is a constant on the stretched horizon, it can be seen that the GHY boundary term reduces to the leading entropy term - it is a constant. First, we note that the classical solution for the dilaton in the bulk from (\ref{a3a}) is $\varphi_+ + k/4 ~(\rho^2)$ when $V(\varphi) = k$. When $V(\varphi)$ is more general, we can still have the dilaton being $\varphi \sim \varphi_+ + C~ (\rho^2) + ....$ for a large class of potentials and the ellipses refer to terms with higher powers of $\rho$. We show this in Appendix A (\ref{A}). We will only consider such potentials in this paper. This class of potentials includes the dimensionally reduced gravity on the sphere.

In subsequent sections, where we consider the gravity path integral, we will only consider variations of the dilaton where the dilaton is held fixed at the boundary. We will consider the dilaton to be $\varphi_{cl} + \tilde \varphi$ where $\varphi_{cl}$ is the classical solution for the dilaton. We will fix the \emph{boundary value} of the dilaton to be the limiting value of the classical solution at the stretched horizon boundary so $\tilde \varphi = 0$ on the boundary. This boundary value is
$\varphi_{b} = \varphi_+ + C ~(\rho^2) + ...$ evaluated at the stretched horizon. Note that if the potential $V(\varphi)=0$, then the boundary value of the dilaton is a constant $\varphi_+ $.

The relevant boundary term to be added with Dirichlet boundary conditions on the induced metric and $\varphi$  to make the variation of the action well-defined is the GHY boundary term. We denote these boundary conditions $DD$ as both the dilaton and induced metric obey Dirichlet boundary conditions.
The GHY boundary term, to leading order in $\varepsilon$ is
\begin{align}
S_{DD}
   = \frac{1}{8\pi G} \frac{2\pi}{\beta}\int_\Delta \!d\tau \,
   \varphi\left(1 - \frac{\beta^2}{4\pi^{2}} \left [ \frac{1}{2}\frac{\tau^{\prime\prime\,2}}{\tau^{\prime\,4}} +\frac{1}{\tau^{\prime\,2}}\{\tau,\sigma\} \right ] \right) \, .
\label{a7a}
\end{align}
Here, $\{\tau,\sigma\}$ is the Schwarzian derivative of $\tau$, defined as
\begin{align}
\{\tau,\sigma\} =  \frac{\tau'''}{\tau'} - \frac{3}{2}\frac{\tau''^{2}}{\tau'^{2}}
\label{a9}
\end{align}
and primes refer to derivatives with respect to $\sigma$.

For $\varphi$ constant, (i.e., when $V(\varphi) = 0$) the boundary GHY action reduces to a constant. Therefore, for interesting boundary dynamics, we will have to choose a potential $V(\varphi)$ which is not zero. If we consider a general potential $V(\varphi)$ with classical solutions such that $\varphi \sim \varphi_+ + C~ (\rho^2) + ....$ then the form of the boundary action (\ref{a7a}) is \emph{universal} at leading order in $\rho$.
The GHY boundary term (\ref{a7a}) is obtained in \cite{carlip} from the leading contribution to the extrinsic curvature coming from the flat metric (\ref{a5a}). For spherically reduced gravity, which is what is relevant for the non-extremal black hole, the next order correction to the extrinsic curvature coming from the exact classical solution to dilaton gravity, the Schwarzschild metric, comes at order $\varepsilon^3$ as we show in Appendix B (\ref{B}). The exact nature of the correction is such that it does not affect the action evaluated at the classical solution nor the one-loop corrections at the order we are considering.
We also observe that for the $\varphi_b$ that we have chosen, \emph{to leading order}, it obeys the differential equation
\begin{align}
\varphi_b = \varphi_+ +  \frac{1}{2}\rho\partial_\rho\varphi \, .
\label{a7}
\end{align}

If instead, in the variation of the action, $\varphi$ and  $n^a\nabla_a\varphi$ are kept fixed at the boundary (we denote this $DN$, for Dirichlet-Neumann), this necessitates adding another boundary term to the action in addition to the GHY term. The total boundary term, evaluated at the stretched horizon is (to leading order in $\varepsilon$, and upon using (\ref{a7}))
\begin{multline}
S_{DN} = \frac{1}{8\pi G}\frac{2\pi}{\beta} \int_{\Delta} \!d\tau\, \varphi_+
   - \frac{1}{16\pi G}\int_\Delta \!d\sigma\, \frac{\beta^2}{4 \pi^{2}}\frac{\varepsilon}{\tau^{\prime\,2}}\partial_\rho\varphi
   \left( \frac{4\pi^2}{\beta^2} \tau^{\prime\,2} - \{\tau,\sigma\}\right) \, .
 \label{a8}
\end{multline}

We have assumed that the boundary value of $\varphi$ is its classical value $\varphi_{b} = \varphi_+ + C\rho^2 +...$. Then,
$\partial_\rho\varphi = 2C\rho$. Alternately, we can say that $\partial_\rho\varphi$ is fixed to be that for the classical solution. Plugging this back in (\ref{a8}) and using the fact that at leading order, $\varepsilon \sim \rho \tau'$, we get
\begin{eqnarray}
S_{DN} = \frac{1}{8\pi G}\frac{2\pi}{\beta} \int_{\Delta} \!d\tau\, \varphi_+
 - \frac{2C}{16\pi G}\int_\Delta \!d\sigma\, \frac{\beta^3}{8\pi^{3}} \frac{\varepsilon^2}{\tau^{\prime\,3}}
   \left( \frac{4\pi^2}{\beta^2} \tau^{\prime\,2} - \{\tau,\sigma\}\right) \, .
 \label{a10}
\end{eqnarray}

Thus, we see that the boundary action at the stretched horizon has the Schwarzian with a pre-factor $\frac{1}{\tau'^{3}}$.
The Schwarzian $\{\tau,\sigma\}$ has an $SL(2,R)$ symmetry - it is invariant under $\tau \rightarrow \frac{a\tau + b}{c\tau + d}$ where
$ad-bc=1$. However, we see that the pre-factor breaks this symmetry both above in (\ref{a10}) and in the GHY boundary term (\ref{a7a}) relevant for the situation when we fix the boundary metric. The action is now only invariant under $\tau \rightarrow \tau + c$ where $c$ is a constant.

\section{Quantizing the boundary action}

One could now consider quantizing the theory given by the boundary action (\ref{a7a}) or (\ref{a8}) in the spirit of similar work in JT gravity. We motivate such a computation further. Consider the bulk partition function in dilaton gravity, where we set Dirichlet boundary conditions for the metric and the dilaton at the boundary,
\begin{equation}
\int [Dg_{ab}]\int [D \varphi] D[\tau] e^{\int d^2x \sqrt{g}(\varphi R + V(\varphi) )} e^{S_{GHY}}
\label{c0}
\end{equation}
Here we have explicitly introduced an integral over the boundary degrees of freedom, which are all configurations corresponding to time reparametrizations of the stretched horizon. This is similar in spirit to what has been done for near-extremal black holes. This is an integral over boundaries, keeping boundary length fixed.
The case when potential $V(\varphi) = c \varphi $ was studied in JT gravity. There, the path integral was simplified by considering the contour for the dilaton integral to go on the imaginary axis \cite{saad}. Then, the integral over the dilaton resulted in a delta function of the factor multiplying the dilaton in the action. The justification given for this contour procedure was that in the corresponding Lorentzian path integral, these configurations were the dominant ones \cite{iliesiu}. This important argument reduced the integral over the metrics to be a sum of constant curvature metrics. One was then left with just the boundary degrees of freedom, i.e., the theory with a Schwarzian action. The dilaton integral resulted in $\delta(R + 1/L^2)$ in the path integral, restricting the integral over metrics to be a sum over constant negative curvature configurations. The case $V(\varphi ) = k$ is similar to the JT gravity case, except now the integral over metrics reduces to a sum over zero curvature configurations.

However, we are interested in more general $V(\varphi)$ such as that in spherically symmetric reduced gravity. The integral over the dilaton and the metric are no longer trivial as in JT gravity. We now use the background field method. We consider the one-loop corrections about the classical solution for the metric and dilaton in the bulk. So,
$g_{ij} = g_{ij}^{cl} + \tilde g_{ij}$ and $\varphi = \varphi_{cl} + \tilde \varphi$ where the fluctuations in the metric and the dilaton vanish at the boundary. Plugging this in the path integral (\ref{c0}), we see that the leading contribution from the boundary degrees of freedom comes from the boundary path integral with classical values for the metric and dilaton,
\begin{equation}
\int D[\tau] e^{S_{GHY} (g^{cl}, \varphi_{cl})}.
\label{c1}
\end{equation}
The near-horizon form of $g^{cl}$ is the flat metric to leading order in $\varepsilon$.
Thus, $S_{GHY}$ is the boundary action of Carlip corresponding to $DD$ boundary conditions, (\ref{a7a}). We confirm this in Appendix B where we show that the correction to the extrinsic curvature from taking the corrections to the flat metric are higher order in $\varepsilon$ than those considered in this paper.

There is also a subtle difference between the boundary path integral (\ref{c1}) and what has been done in JT gravity for near-extremal black holes \cite{maldacena} which we spell out clearly to avoid confusion.
In the case of near-extremal black holes, the integral over the dilaton in the bulk could be done exactly. The integral over the bulk metric restricted the metric to have constant negative curvature. To compute $S_{GHY}$, we need as input both the dilaton on the boundary as well as the ambient metric in the bulk. Thus the ambient metric in the boundary action was taken to be the anti-de Sitter metric. The boundary value of the dilaton was an external coupling and the corresponding boundary path integral was also done exactly in the case when the dilaton was a constant.

Here, our bulk one-loop computation implies that our boundary value of the dilaton is what it is for the classical solution. This does depend on $\rho(\sigma)$ and thus changes as we change the boundary. This seems to be the only way to find the leading contribution from the integral over boundaries in the case when we cannot perform the bulk path integral exactly and have to resort to one-loop approximations about the classical solution. We make some comments on viewing the dilaton as an external coupling in section IX.
We will get a similar result with the path integral with $DN$ boundary conditions with $\varphi$ and $n^a\nabla_a\varphi$ kept fixed at the boundary with the boundary action $S_{DN}$ given by (\ref{a8}), except that in the end, we will also have to integrate over the length of the stretched horizon boundary, since it is not held fixed.

With these motivations, we now quantize the boundary actions (\ref{a7a}) and (\ref{a8}). In particular, we will attempt to compute the one-loop corrections to the classical solution for the time reparametrization in both cases. For this, we first need the appropriate measure for integrating over $\tau(\sigma)$. We show in the next section that the boundary path integral is an integral on $Diff(S^1)/S^1$ which is a symplectic manifold. It is in fact an ordinary coadjoint orbit of the Virasoro group (basics of these orbits are reviewed in the next section). There is a corresponding symplectic form with an associated symplectic measure which we can use. This measure depends on two numbers: the particular coadjoint vector whose orbit we are considering and the central charge.
\section{Reparametrizations of the stretched horizon}

Treating the stretched horizon as a boundary, as we have seen, the gravitational action is not invariant under diffeomorphisms that change the boundary. As $\varepsilon$ is a constant on the boundary, $\rho$ and $\tau$ are not both independent. Further, $\rho \sim O(\varepsilon)$. Therefore, at leading order in $\varepsilon$, $\tau(\sigma)$ represents reparametrizations of the stretched horizon, and the boundary action explicitly depends on it. The boundary action is invariant under a constant shift in $\tau$ which we can take as a rigid rotation since $\tau$ is periodic. Thus, we can take $\tau (\sigma)$ as an element of
$Diff(S^1)/S^1$ as for example in JT gravity where the time reparametrization is an element of $Diff(S^1)/SL(2,R)$ \cite{stanfordwitten}. In \cite{stanfordwitten}, $Diff(S^1)/SL(2,R)$ was the exceptional orbit of a certain coadjoint vector under the action of the Virasoro group. Similarly $Diff(S^1)/S^1$ is the ordinary orbit of a certain set of coadjoint vectors under the action of the Virasoro group \cite{wittencoadjoint}, \cite{alekseev}.
Let us briefly review the salient features of coadjoint vectors. Consider the Lie group $Diff(S^1)$. Its elements are monotonic functions $\phi(\theta + 2\pi) = \phi(\theta) + 2\pi,~~ \theta \in [0,2\pi)$. The Lie algebra is given by infinitesimal diffeomorphisms which are vector fields on $S^1$, given by $f(\theta) \frac{d}{d\theta}$. Denote by
$\hat Diff(S^1)$ the Virasoro group, which is the central extension of $Diff(S^1)$. Its Lie algebra is given by elements $f\frac{d}{d\theta} -ia_f c$ denoted by $(f, a_f)$ where $a_f$ is a real number multiplying the central element $c$. The commutator of two elements of this algebra is defined as (we follow the conventions of \cite{cotlerjensen} )
\begin{eqnarray}
[f \frac{d}{d\theta} - i a_f c, g\frac{d}{d\theta}-ia_g c] = (fg' - gf')\frac{d}{d\theta} + \frac{ic}{48\pi}\int_{0}^{2\pi}d\theta (f g''' - g f''')
\label{b0a}
\end{eqnarray}
These are the adjoint vectors. Coadjoint vectors are defined to be linear functionals acting on adjoint vectors. For the Virasoro group, they are given by $b(\theta)d\theta^2 + i t\tilde c$, denoted $(b,t)$ where $\tilde c$ is the central element of the coadjoint representation and $t$ is a real number.
The action of a coadjoint vector $(b,t)$ on the adjoint vector $(f,a)$ is given by the pairing
\begin{eqnarray}
<(b,t), (f,a)> = \int_{0}^{2\pi} d\theta bf ~+~at.
\label{b0b}
\end{eqnarray}
The orbit of a coadjoint vector under the Virasoro group is given by how it changes under a finite diffeomorphism.  Let the inverse diffeomorphism be $\phi(\theta)$.
Then the coadjoint vector transformation is
\begin{equation}
(b_{0}, c) \rightarrow (b(\phi), c) = (b_0 \phi'^{2} - \frac{c \{\phi, \theta \}}{24\pi}, c).
\label{b0c}
\end{equation}

Here, $\{\phi, \theta \}$ is the Schwarzian derivative of $\phi$. For $b_0 \neq \frac{-cn^2}{48\pi}$ where $n$ is an integer, we have the ordinary coadjoint orbits of the Virasoro group, $Diff(S^1)/S^1$. The other orbits are referred to as exceptional orbits. The first exceptional orbit $b_0 = - \frac{c}{48\pi}$ is the manifold $Diff(S^1)/SL(2,R)$ .
In JT gravity, the time reparametrization is an element of $Diff(S^1)/SL(2,R)$ as the Schwarzian action is invariant under this symmetry. However, in the gravitational boundary action, which is a Schwarzian multiplied by a pre-factor, the pre-factor is not invariant under $SL(2,R)$ transformations. So the diffeomorphism can actually be viewed as an element of the ordinary coadjoint orbit for $b_0 \neq \frac{-cn^2}{48\pi}$.

The fact that $Diff(S^1)/SL(2,R)$ is a coadjoint orbit under the action of the Virasoro group was used by Stanford and Witten to prove that the partition function of the Schwarzian theory is one-loop exact. Coadjoint orbits are symplectic, and one can define a symplectic form on them.
The symplectic form is defined as follows: consider the adjoint vectors $F_1 = (f_1 (\theta), a_1)$ and $F_2 = (f_2 (\theta), a_2)$. Consider the coadjoint vector $(b,c)$. Let $X_1$ and $X_2$ be elements of the tangent space to the orbit at $b$ obtained from diffeomorphisms $F_1$ and $F_2$ (the detailed action of the diffeomorphism on $b$ can be found in \cite{wittencoadjoint}). Then, the symplectic form $\omega$ on the coadjoint orbit is given by
\begin{equation}
\omega(X_1, X_2) = - <(b, c), [F_1, F_2]>
\label{b0d}
\end{equation}

Following Stanford and Witten, we will consider $\sigma(\tau)$ to label points on this coadjoint orbit, and write the symplectic form in terms of the inverse diffeomorphism $\tau(\sigma)$.
The symplectic form on the coadjoint orbit $Diff(S^1)/S^1$ of a coadjoint vector $b_0$  is
\begin{eqnarray}
\omega = - \int_{0}^{2\pi} d\sigma [ \frac{c}{48 \pi} \frac{d\tau' \wedge d\tau''}{\tau'^{2}} + \frac{4\pi^{2}}{\beta^2} b_{0} d\tau \wedge d\tau'] .
\label{b1}
\end{eqnarray}
Here, $c$ is the central charge of the Virasoro algebra.
The Pfaffian of $\omega$ is the natural measure on a symplectic manifold. Our strategy is to compute the `boundary path integral' over $\tau(\sigma)$ with this measure and with Carlip's boundary actions. The measure depends on the constants $c$ and $b_0$. Before we move on to the next section, we also revisit the result in \cite{stanfordwitten} that the Schwarzian theory is one-loop exact.

As is done in \cite{stanfordwitten}, let us consider an infinitesimal $Diff (S^1)$ transformation $\delta \sigma = \alpha (\sigma)$ under which we have
$\delta \tau = \alpha (\sigma) \tau'$. This is a vector field $V_{\alpha}$ on the space of functions $\tau$. On the space of functions $Diff(S^1)/S^1$, we can consider the interior product $\iota_{V_\alpha} \omega$. The task then is to find a function (the Hamiltonian) $H_{\alpha}$ such that $\iota_{V_\alpha} \omega = dH_{\alpha}$. Stanford and Witten argue that the integral
\begin{eqnarray}
\int [d\tau] Pf(\omega) e^{(H_{\alpha}/g^2)}
\label{b2}
\end{eqnarray}
is one-loop exact by the Duistermaat-Heckman formula \cite{dhformula}. Here $Pf(\omega)$ is the Pfaffian of $\omega$. It turns out that for the space of functions $Diff(S^1)/SL(2,R)$ which is the coadjoint orbit of a special coadjoint vector, the Hamiltonian is precisely the Schwarzian action. The result implies that the path integral with the Pfaffian measure and the Schwarzian action is one-loop exact. Stanford and Witten also give a proof based on fermionic localization for this result.

To investigate whether we can obtain such a result in our case,
we need to find the $H_{\alpha}$ corresponding to the symplectic form (\ref{b1}). We would like to see if the boundary actions (\ref{a7a}) or (\ref{a8}) are of the form $H_{\alpha}$ for some $\alpha$. We have not found a ready expression for $H_{\alpha}$ corresponding to the ordinary orbits. So we compute it in Appendix C (\ref{C}).
Here, we merely give the final answer.
\begin{eqnarray}
H_{\alpha} = \int_{0}^{2\pi} d\sigma \alpha [ \frac{2c}{48\pi}\{\tau,\sigma\} - \frac{4\pi^2}{\beta^2} b_0 \tau'^2 ].
\label{b12}
\end{eqnarray}
We note that this is not of the form of Carlip's boundary action (\ref{a7a}) or (\ref{a8}) for any $b_0$ and $c$ despite the fact that $\tau(\sigma)$ can actually be viewed as an element of the ordinary coadjoint orbit for $b_0 \neq \frac{-cn^2}{48\pi}$. However, the one-loop Carlip boundary action (\ref{a8}) quadratic in fluctuations about the classical solution is equal to (\ref{b12}) evaluated about the classical solution to quadratic order in fluctuations for a particular ratio $\frac{b_0}{c}$ in the $DD$ case, and a different numerical constant for the ratio in the $DN$ case. It would be interesting to see if this implies that the higher loop corrections from the fluctuation are zero. We may choose to fix the ratio $\frac{b_0}{c}$ using this, however many of our conclusions in the paper do not depend on the exact numerical value of the constant $\frac{b_0}{c}$.

\section{The classical contribution to the boundary path integral }
Let us consider the boundary action of Carlip for $DN$ boundary conditions, where we fix $\varphi$ and $n^a\nabla_a\varphi$ on the boundary. This can be written as
\begin{equation}\label{5}
S_{DN}=\frac{\varphi_+}{4G}+ S^{DNextra}
\end{equation}
where
\begin{equation}\label{6A'}
S^{DN extra} =  - \frac{1}{16\pi G}\int_{\Delta} d\sigma \frac{\varepsilon}{\tau'^2}\partial_{\rho}\varphi \Big(\tau'^2-\frac{\beta^2}{4\pi^2}\{\tau,\sigma\}\Big).
\end{equation}
For Dirichlet  boundary condition on $\varphi$ and on the metric we get
\begin{equation}\label{6A''}
S^{DDextra} =\frac{\ell}{32\pi^2 G}\int_{\Delta} d\sigma \partial_{\rho}\varphi\Big(1-\frac{\beta^2}{(2\pi)^2}\Big(\frac{\tau''^2}{2\tau'^2} + \frac{1}{\tau'^2}\{\tau,\sigma\}\Big)\Big)
\end{equation}
where $\varphi_{+}$ is the value of $\varphi$ at $\rho =0$. At leading order in $\rho$, we have $\varphi=\varphi_{+} +\rho\partial_{\rho}\varphi $ or $\varphi = \varphi_{+} + C \rho^2 + ...$. Using $\rho=\varepsilon \frac{\beta}{2\pi}/ \tau'$ we get
\begin{equation}\label{6A}
S^{DNextra} =  - \frac{1}{g^2}\int_{\Delta} d\sigma \frac{\beta }{2\pi\tau'^3}\Big(\tau'^2-\frac{\beta^2}{4\pi^2}\{\tau,\sigma\}\Big)
\end{equation}
where the coupling constant in the action  $\frac{1}{g^2}=\frac{2C\ell^2}{64\pi^3 G}$. We can write the action for $DD$ case as,
\begin{equation}\label{6Ab}
S^{DDextra}=\frac{1}{g^2}\frac{\beta}{2\pi} \int_{\Delta} d\sigma  \frac{1}{\tau'}\Big(1-\frac{\beta^2}{(2\pi)^2}\Big(\frac{\tau''^2}{2\tau'^2} + \frac{1}{\tau'^2}\{\tau,\sigma\}\Big)\Big)
\end{equation}
 The classical equation of motion (EOM) for $\tau$ in the boundary action $S^{DNextra}$ is
\begin{equation}\label{7}
\Big(-\frac{1}{\tau'^2}+\frac{\beta^2}{4\pi^2}\Big(\frac{4}{\tau'^4} \{\tau,\sigma \} -\frac{3\tau''^2}{2\tau'^6}\Big)\Big)'=0
\end{equation}
It can easily be seen that $\tau(\sigma) = \frac{\beta \sigma}{2\pi} $ is a solution to the above EOM. Actually the classical solution can be any integer multiple of this solution. This will correspond to multiple windings which we do not consider here. We can also choose the reparametrization with the opposite orientation, which we will not do. We will have more to say about this solution later. We now compute the action evaluated at this classical solution. Only the first two terms in the action $S_{boundary}$ ( \ref{5}) will contribute.
In the $DN$ case, the corresponding contribution from the action evaluated at the classical solution to the boundary path integral is denoted \footnote{With Carlip's sign convention for the action, the partition function is defined as $\int D[\phi]e^{S[\phi]}$. This is determined from Carlip's action as the choice of sign yielding the correct Bekenstein-Hawking entropy.}
$Z^{DN}_{cls}$ and we have
\begin{equation}\label{8}
log Z^{DN}_{cls} =S_{DN}^{cls}=\frac{\varphi_+}{4G} -\frac{2C\ell^2}{32\pi^2 G }
\end{equation}
 This expression is obtained by substituting the value of the coupling constant. Similarly, we can compute the EOM for $DD$ boundary conditions and it can easily be checked that $\tau(\sigma) =\frac{\beta \sigma}{2\pi} $ is again the classical solution. Therefore, the contribution from the classical solution in the boundary path integral in the $DD$ case is $Z^{DD}_{cls}$, given by
\begin{equation}\label{8A}
\log{Z^{DD}_{cls}}=S_{DD}^{cls}=\frac{\varphi_+}{4G} +\frac{2C\ell^2}{32\pi^2 G }
\end{equation}
This is as expected, $\frac{\varphi|_{\Delta}}{4G}$.

\section{Perturbations around the classical solution}\label{s5}
The perturbation around the classical solution is
\begin{equation}\label{9}
\tau(\sigma)= \frac{\beta}{2\pi}\sigma + \lambda\epsilon(\sigma)
\end{equation}
where $\epsilon(\sigma)$ is a perturbation about the classical solution and $\lambda$ is a perturbative parameter, which is of order of $ O(\varepsilon)$. For example one can choose\footnote{We will see that partition function will be independent of the choice of $\lambda$. } $\lambda=\frac{\sqrt{G} \ell}{\beta}$, also notice that $\epsilon(\sigma + 2\pi)= \epsilon(\sigma)$. Expanding the action (\ref{5}) up to quadratic order in $\epsilon$ gives,
\begin{equation}\label{11}
S^{DNextra}=  -\frac{1}{g^2}\int_{\Delta}\Big\{1-\frac{2\pi \lambda}{\beta} \epsilon'+\frac{4\pi^2\lambda^2}{\beta^2}\epsilon'^2+\frac{10\pi^2}{\beta^2}\lambda^2\epsilon'''\epsilon'\Big\}d\sigma .
\end{equation}
We can write $S^{DNextra}= -2C \ell^2/ 32\pi^2G + S^{DN}_{\epsilon}$ , where we got the first term by substituting the value of the coupling and
\begin{equation}\label{12}
S^{DN}_\epsilon= -\frac{4\pi^2 \lambda^2}{g^2 \beta^2}\int_{\Delta}d\sigma \Big(\epsilon'^2+\frac{5}{2}\epsilon''' \epsilon'\Big) .
\end{equation}
Now using the fact that $\epsilon$ is periodic in $2\pi$, we can expand it in a Fourier basis.
\begin{equation}\label{13}
\epsilon(\sigma) = \sum_{n=-\infty}^{\infty} C_n \exp(in\sigma).
\end{equation}
\nocite{*}
Plugging this expansion in (\ref{12}), we get
\begin{equation}\label{14}
S^{DN}_\epsilon= -\frac{4\pi^2 \lambda^2}{g^2 \beta^2} \int_{\Delta} \Big \{\sum_{n, m} \Big( -nmC_{n}C_{m}+\frac{5}{2}m^3 n C_{n}C_{m}\Big) \exp{i(n+m)\sigma}\Big\} d\sigma .
\end{equation}
Using the identity
\begin{equation}\label{15}
\int_{0}^{2\pi} \exp{i(n+m)\sigma}~ d\sigma = 2\pi \delta_{n+m,0}
\end{equation}
the equation (\ref{14}) becomes
\begin{equation}\label{16}
S^{DN}_\epsilon=\frac{(2\pi)^3}{\beta^2}\frac{\lambda^2}{g^2}\sum_{n=-\infty}^{n=\infty}\Big(\frac{5}{2}n^4-n^2\Big)C_{n}C_{-n}
\end{equation}
We have already seen that $\tau(\sigma)= \beta \sigma /2\pi$ is the classical solution for $S^{DDextra}$. We can perturb the action about this classical solution as in equation (\ref{9}) for  $S^{DNextra}$. Now if we take $\tau(\sigma)= \frac{\beta}{2\pi}\sigma + \lambda \epsilon(\sigma)$ and expand the action (\ref{6A''}) to quadratic order in $\epsilon$, we will get
\begin{eqnarray}\label{16B}
S^{DDextra} &=& \frac{2C\ell^2}{32 \pi^2 G} + S^{DD}_{\epsilon} \nonumber \\
&=& \frac{2C\ell^2}{32 \pi^2 G} + \frac{4\pi^2 \lambda^2}{g^2\beta^2}\int_{\Delta}\Big(\epsilon'^2- 2\epsilon''' \epsilon'\Big) d\sigma .
\end{eqnarray}
We can again write $\epsilon(\sigma) = \sum_{n=-\infty}^{\infty} C_n \exp(in\sigma)$, which will give
\begin{equation}\label{16A}
S^{DD}_{\epsilon}=\frac{(2\pi)^3\lambda^2}{g^2\beta^2}\sum_{n}\Big(2n^4 +n^2\Big) C_{n}C_{-n}
\end{equation}
Since $\epsilon$ is real, from (\ref{13}) we get
\begin{equation}\label{17}
C_{-n}= \bar{C}_{n}
\end{equation}
where $ \bar{C}_{n}$ is the complex conjugate of $C_{n}$.
\section{Calculation of 1-loop partition function}\label{s6}
Let us now compute the boundary path integral at one loop about the classical solution in the $DN$ case. We use the fact that the path integral is over $Diff(S^1)/S^1$ which has the symplectic form (\ref{b1}) whose Pfaffian will give the symplectic measure. $c$ is central charge and $b_0$ is some constant coadjoint vector.
The one loop partition function for the $DN$ case can written as,
 \begin{equation}\label{25}
 Z^{DN}_{1-loop}=Z^{DN}_{cls} \int \prod_{n=1}^{\infty}\lambda^2 dC_n d\bar{C}_n Pf(\omega) \exp{\Big(\frac{(2\pi)^3\lambda^2}{g^2\beta^2}\sum_{n=-\infty}^{n=\infty}\Big(\frac{5}{2}n^4-n^2\Big)C_{n}C_{-n} \Big)}
 \end{equation}
The actual partition function for this case involves one more integral over the length of the stretched horizon, which is not fixed for these choices of boundary conditions.
To get the measure we just have to calculate the Pfaffian of the symplectic form. One can expand about the identity in $Diff(S^1)$ i.e ($\tau=\frac{\beta}{2\pi} \sigma+\lambda\sum_{n=-\infty}^{\infty} C_n \exp(in\sigma)$). Calculation of $\omega$ about the classical solution ( $C_n=0$) will give,
\begin{equation}\label{27}
\omega= i\frac{4\pi^2 \lambda^2}{\beta^2}\sum_{n}\frac{c}{24}\Big(n^3+\frac{48\pi b_0 n}{c}\Big) dC_n \wedge dC_{-n} .
\end{equation}
where $Pf(\omega)$ for the above symplectic form is
\begin{equation}\label{28}
Pf(\omega)=\prod_{n=1}^{\infty}\frac{4\pi^2}{\beta^2} \Big(\frac{ic}{24}\Big) \Big(n^3+\frac{48\pi b_0 n}{c}\Big) .
\end{equation}
$n=0$ is a zero mode that we just omit from the measure as the path integral is on $Diff(S^1)/S^1$.
We can write the above equation as
\begin{equation}\label{29}
Pf(\omega) =Pf(\omega)|_{\beta=2\pi}\prod_{n=1}^{\infty}\frac{4\pi^2}{\beta^2} =\frac{\beta}{2\pi} (Pf(\omega)|_{\beta=2\pi} ).
\end{equation}
After  regularization,  $Pf(\omega)|_{\beta=2\pi}$ will give some quantity which depends only on $c$ and $b_0$. This quantity is derived in appendix E (\ref{E}). Now, let us compute the partition function with the symplectic measure
\begin{equation}\label{18}
Z^{DN}_{1-loop}=\frac{ Z^{DN}_{cls}\beta}{2\pi} \int (Pf(\omega)|_{\beta=2\pi} )\prod_{n=1}^{\infty} \lambda^2 dC_n d\bar{C}_n \exp{\Big(\frac{(2\pi)^3\lambda^2}{g^2\beta^2}\sum_{n=-\infty}^{n=\infty}\Big(\frac{5}{2}n^4-n^2\Big)C_{n}C_{-n} \Big)}
\end{equation}
 We can write $C_{n} \bar{C}_{n}$ as $(C^{Re}_n)^2+(C^{Im}_n)^2$ and $dC_n d\bar{C}_n $ as $-2 i d(C^{Re}_n) d(C^{Im}_n) $. Let $\alpha_{DN}=\frac{ Z^{DN}_{cls}\beta}{2\pi}(Pf(\omega)|_{\beta=2\pi}  $. Then the partition function becomes
\begin{equation}\label{19}
Z^{DN}_{1-loop}=\alpha_{DN} \int \prod_{n=1}^{\infty}(-2i\lambda^2 )d(C^{Re}_n) d(C^{Im}_n)  \exp{\Big(\frac{16\pi^3\lambda^2}{g^2\beta^2}\sum_{n=1}^{n=\infty}\Big(\frac{5}{2}n^4-n^2\Big)\Big[(C^{Re}_n)^2+(C^{Im}_n)^2\Big] \Big)}
\end{equation}

We see from (\ref{19}) that we have `Gaussians with the wrong sign', and a divergent path integral \footnote{It was suggested to us by Steve Carlip that we could investigate instead, the action for $\sigma(\tau)$ to see if we get finite one-loop fluctuations about the classical solution. We have investigated this, but the problem persists with the action for $\sigma(\tau)$.}. There are two possibilities: either we are dealing with a unstable classical solution or with an indefinite action reminiscent of the conformal mode problem in quantum gravity \cite{GHP}. In Appendix D (\ref{D}), we compute the one-loop action about a general classical solution in the $DN$ case (the $DD$ case is similar). The sign of the one-loop part of the action seems to depend non-trivially on the classical solution. But in that Appendix, if the derivative of the classical solution $\tau'_{c} > 0$ and the fluctuation $\delta \tau$ can be expanded in Fourier modes, then the one-loop action has the wrong sign. One trivial way that we can change the wrong signs is by choosing the classical solution $\tau =- \frac{\beta}{2\pi} \sigma$, corresponding to changing the orientation of the $S^1$. But if we do this, and restrict to negative orientation $\tau$, the leading term in the partition function with the classical action will change sign and the entropy will pick up a minus sign. In order to fix that negative sign, we will have to change the convention for the path integral to $\int D[\phi]e^{- S[\phi]}$. This will lead to the correct sign for the entropy but our Gaussians in the one loop computation will again pick up the wrong sign corresponding to opposite orientation $\tau (\sigma)$.

If the classical solution is unstable, it signals that we cannot ascribe boundary degrees of freedom to the stretched horizon and obtain a finite partition function as happens for the near-extremal black hole in JT gravity. However, the problem we are facing could be due to the indefiniteness of the action as well. Unlike the Schwarzian action in the case of near-extremal black holes, the Carlip actions are not $SL(2,R)$ invariant and there is no scaling symmetry, so whether the orientation is positive or negative also seems to affect the sign. If the problem has arisen due to the indefiniteness of this action, then one can do an analytic continuation
similar to Gibbons, Hawking and Perry \cite{GHP} --- i.e., we can take the perturbations lying in this space to be purely imaginary in order to make the path integral converge in the one-loop approximation. This changes the relative sign between the one loop correction and the classical action. Since such a computation is doable, let us go ahead, do this and see what answer we get.

Therefore to make the above path integral convergent, we can take $C^{Re}_n\rightarrow i C^{Re}_n$ and $C^{Im}_n\rightarrow i C^{Im}_n$. This yields
\begin{equation}\label{19A}
Z^{DN}_{1-loop}=\alpha_{DN} \int \prod_{n=1}^{\infty}(2i\lambda^2 )d(C^{Re}_n) d(C^{Im}_n)  \exp{\Big(-\frac{16\pi^3\lambda^2}{g^2\beta^2}\sum_{n=1}^{n=\infty}\Big(\frac{5}{2}n^4-n^2\Big)\Big[(C^{Re}_n)^2+(C^{Im}_n)^2\Big] \Big)}
\end{equation}
The above equation can be thought of as a product of two Gaussian integrals for each $n$. Also, note that $\lambda^2$ will cancel out. Using standard Gaussian integrals, we get
\begin{equation}\label{20}
Z^{DN}_{1-loop}=\frac{ Z^{DN}_{cls}\beta}{2\pi}(Pf(\omega)|_{\beta=2\pi} )\prod_{n=1}(2i\lambda^2) \prod_{n=1}^{\infty} \Big\{\frac{ g^2 \beta^2}{16\pi^2 \lambda^2 \Big(\frac{5}{2}n^4-n^2\Big)}\Big\}
\end{equation}
Using zeta regularized product identities\footnote{$\prod_{n=1}^{\infty} \lambda_n= \exp-\zeta_\lambda'(0) $ where $\zeta_\lambda(s)=\sum_{n=1}^{\infty} \lambda_n^{-s}$.}(for details look at appendix E \ref{E}), we get
 \begin{equation}\label{21}
 Z^{DN}_{1-loop}=\frac{ Z^{DN}_{cls}\beta}{2\pi} \mu \exp{\Big(-\log\Big(\frac{16\pi^2}{g^2 \beta^2}\Big)\zeta(0) -\sum_{n=1}^{\infty}\log{\Big(\frac{5}{2}n^4-n^2\Big)}\Big)}
 \end{equation}
 where,
 \begin{equation}\label{21A}
 \mu= (Pf(\omega)|_{\beta=2\pi} )\prod_{n=1}(2i)
 \end{equation}

 \begin{equation}\label{22}
 Z^{DN}_{1-loop}=\#_{DN} \frac{ Z^{DN}_{cls}\beta}{2\pi}\mu\sqrt{\frac{16\pi^2}{g^2 \beta^2}}.
 \end{equation}
The value $\#_{DN}$ and $\mu$ is calculated in appendix E (\ref{E}) using zeta function regularization. We can put the value of $g^{-2}$ to make the $\ell$ dependence explicit,
 \begin{equation}\label{23}
 Z^{DN}_{1-loop}=\#_{DN} \frac{ Z_{cls}\beta}{2\pi} \mu \sqrt{\frac{2C\ell^2}{4\pi G \beta^2}}
 \end{equation}
 If we substitute the value of $\mu$, computed in appendix E ({\ref{E}), we get the full partition function as \footnote{Depending on sign of $b_0$,the $\mu$ have either a hyperbolic sine or sine. We have written $\mu$ assuming $b_0>0$, in the case $b_0<0$ replace $\sinh$ with $\sin$.},
 \begin{equation}\label{23A}
 Z^{DN}_{1-loop}= \frac{ Z_{cls}\beta}{2\pi} \Big\{\frac{1}{4\pi\sin{\frac{\sqrt{2}\pi}{\sqrt{5}}}}\Big\} \Big(i\sqrt{\frac{2}{b_0}}\sinh{\Big(\pi \sqrt{\frac{48\pi b_0}{c}}\Big)}\Big)\sqrt{\frac{2C\ell^2}{4\pi G \beta^2}}
 \end{equation}.

 Finally, we have to do an integral over $\ell$, the length of the stretched horizon, which is not fixed for $DN$ boundary conditions. However, we will not bother to do this, since we see that terms involving $\beta$ cancel out in the partition function, and it is independent of $\beta$. It therefore does not contribute to thermodynamics.
 This is true whatever be the numerical value of the ratio $\frac{b_0}{c}$ --- even with this ratio fixed, the partition function is only obtained up to an overall constant factor, but it seems to be $\beta$ independent whatever be the numerical values of these constants.
Following the same steps, we can write the one loop partition function for $DD$ boundary conditions,
 \begin{equation}\label{31B}
 Z^{DD}_{1-loop}=Z^{DD}_{cls} \int \prod_{n=1}^{\infty} \lambda^2 dC_n d\bar{C}_n Pf(\omega) \exp{\Big(\frac{(2\pi)^3\lambda^2}{g^2 \beta^2}\sum_{n}\Big(2n^4 +n^2\Big) C_{n}C_{-n}\Big)}.
 \end{equation}
We have already computed the Pfaffian in (\ref{29}). We can write the above equation as
 \begin{equation}\label{31C}
 Z^{DD}_{1-loop}=\frac{ Z^{DD}_{cls}\beta}{2\pi} (Pf(\omega)|_{\beta=2\pi} )\int \prod_{n=1}^{\infty}\lambda^2 dC_n d\bar{C}_n \exp{\Big(\frac{(2\pi)^3 \lambda^2}{g^2 \beta^2}\sum_{n}\Big(2n^4 +n^2\Big) C_{n}C_{-n}\Big)}
 \end{equation}
 Let $\alpha_{DD}=\frac{ Z^{DD}_{cls}\beta}{2\pi}(Pf(\omega)|_{\beta=2\pi}  $. Then the partition function becomes
 \begin{equation}\label{31C1}
Z^{DD}_{1-loop}=\alpha_{DD} \int \prod_{n=1}^{\infty}(-2i\lambda^2 )d(C^{Re}_n) d(C^{Im}_n)  \exp{\Big(\frac{16\pi^3\lambda^2}{g^2\beta^2}\sum_{n=1}^{n=\infty}\Big(2n^4 +n^2\Big)\Big[(C^{Re}_n)^2+(C^{Im}_n)^2\Big] \Big)}.
\end{equation}
The above path integral again is divergent. Similar to the $DN$ case, we can analytically continue $C^{Re}_n\rightarrow i C^{Re}_n$ and $C^{Im}_n\rightarrow i C^{Im}_n$ in order to make the path integral convergent. This yields
\begin{equation}\label{31C2}
Z^{DD}_{1-loop}=\alpha_{DD} \int \prod_{n=1}^{\infty}(2i\lambda^2 )d(C^{Re}_n) d(C^{Im}_n)  \exp{-\Big(\frac{16\pi^3\lambda^2}{g^2\beta^2}\sum_{n=1}^{n=\infty}\Big(2n^4 +n^2\Big)\Big[(C^{Re}_n)^2+(C^{Im}_n)^2\Big] \Big)} .
\end{equation}
The above equation can be thought of as a product of two Gaussian integrals for each $n$. Using standard Gaussian integrals, we get,
 \begin{equation}\label{31D}
 Z^{DD}_{1-loop}=\frac{ Z^{DD}_{cls}\beta}{2\pi}(Pf(\omega)|_{\beta=2\pi} )\prod_{n=1}(2i) \prod_{n=1}^{\infty} \Big\{\frac{ \beta^2 g^2}{32\pi^2}\Big\}\frac{1}{(n^2(n^2+\frac{1}{2}))}
 \end{equation}
Now, using zeta function product identities given in appendix E (\ref{E})
\begin{equation}\label{31A}
Z^{DD}_{1-loop}= \#_{DD}\frac{Z^{DD}_{cls}\beta}{8\pi^3}\mu\sqrt{\frac{32\pi^2}{ \beta^2 g^2}}
\end{equation}
where $\mu$ is defined in the equation (\ref{21A}). Substituting the values of $\mu$ and
$\#_{DD}$ computed in appendix E (\ref{E}), we get \footnote{Depending on sign of $b_0$,the $\mu$ has $\sinh$ or $\sin$. We have written $\mu$ assuming $b_0>0$, in the case $b_0<0$ replace $\sinh$ with $\sin$.}
\begin{equation}\label{31B2}
Z^{DD}_{1-loop}= \frac{Z^{DD}_{cls}}{8\pi^3} \Big\{ \frac{1}{4\sqrt{2}\pi} \frac{1}{\sinh{\frac{\pi}{\sqrt{2}}}} \Big\} \Big(i\sqrt{\frac{2}{b_0}}\sinh{\Big(\pi \sqrt{\frac{48\pi b_0}{c}}\Big)}\Big)\sqrt{\frac{2C\ell^2}{2\pi G }}
\end{equation}
$\#_{DD}$ is a numerical constant evaluated in appendix E (\ref{E}). Notice that in the equation (\ref{23}) and (\ref{31A}) there is no explicit $\beta$ dependence. So, after the analytic continuation of the field, there is no nontrivial thermodynamics obtained from the one-loop partition function for the boundary action with the symplectic measure on $Diff(S^1)/S^1$. Further, it is imaginary due to the analytic continuation. We conclude that there does not seem to be any way of nontrivially associating boundary degrees of freedom for the non-extremal black hole in such a way that it affects thermodynamics at least at one-loop. This is very different from what has been obtained for the near-extremal black hole where the Schwarzian describes boundary degrees of freedom in the near-horizon region and the resulting partition function has a non-trivial $\beta$ dependence. Finally, we would like to add that it was pointed out to us by Steve Carlip that if one was only interested in the $\beta$ dependence, then one could simply scale the Fourier mode $C_n = \beta D_n$ and then the $\beta$ dependence will go away both in the action and measure, so it is easy to see that $Z$ is $\beta$ independent. This happens precisely because of the prefactors multiplying the Schwarzian in the action (which is what is different to the near-extremal case) as well as the choice of the symplectic measure.

\section{Changing the fundamental assumptions of the model}\label{G}
 As we saw, the one-loop computation about the classical solution does not seem to affect thermodynamics. Here we will relax one of the assumptions in Carlip's model. We analyze a general boundary action on the stretched horizon. Instead of assuming $\rho' \sim O(\varepsilon^{2})$, we assume $\rho' \sim O(\epsilon)$.  Consider the action obtained in \cite{carlip}
 \begin{equation}\label{G1}
 I_{bdry}=\frac{1}{8 \pi G}\int_{\Delta}d\sigma \varphi\Big(\tau'-\Big(1-\frac{\rho'^2}{\varepsilon^2}\Big)^{-1/2}\Big(\frac{\rho'}{\varepsilon}\Big)'\Big)
 \end{equation}
As before,
\begin{equation}\label{G2}
\varepsilon^2=\rho'^2+\kappa^2 \tau^2\rho^2 .
\end{equation}
Notice that when $\phi$ is constant, the action is just a total derivative. Also,
one can always choose the parametrization in such a way that $\varepsilon$ is constant. Using (\ref{G2}) we can write the action completely in terms of $\rho$ and then we can consider the action as a functional of $\rho$ instead of $\tau$.
\begin{equation}\label{G3}
I_{bdry}=\frac{1}{8\pi G}\int_{\Delta}\varphi\Big(\frac{1}{\rho}\sqrt{\varepsilon^2-\rho'^2}-\frac{\rho''}{\sqrt{\varepsilon^2-\rho'^2}}\Big)
\end{equation}
The Euler Lagrange equation is
\begin{equation}\label{G4}
\frac{\partial \mathcal{L}}{\partial \rho}-\frac{d}{d\sigma}\frac{\partial \mathcal{L}}{\partial \rho'}+\frac{d^2}{d\sigma^2}\frac{\partial \mathcal{L}}{\partial \rho''}=0
\end{equation}
where $\mathcal{L}$ is Lagrangian for the action in (\ref{G3}).
\begin{equation*}
\frac{\partial \mathcal{L}}{\partial \rho}=\frac{1}{8\pi G}\Big(\partial \varphi\Big(\frac{1}{\rho}\sqrt{\varepsilon^2-\rho'^2}-\frac{\rho''}{\sqrt{\varepsilon^2-\rho'^2}}\Big)-\frac{\varphi}{\rho^2}\sqrt{\varepsilon^2-\rho'^2}\Big)
\end{equation*}
\begin{multline}\label{G5}
\frac{d}{d\sigma}\frac{\partial \mathcal{L}}{\partial \rho'}=\frac{1}{8\pi G}\Big(-\partial_{\rho}\varphi\Big(\frac{\rho'^2}{\rho\sqrt{\varepsilon^2-\rho'^2}}+\frac{\rho'^2 \rho''}{(\varepsilon^2-\rho'^2)^{3/2}}\Big) +\varphi\Big(\frac{\rho'^2}{\rho^2 \sqrt{\varepsilon^2-\rho'^2}}
- \frac{\rho''}{\rho\sqrt{\varepsilon^2-\rho'^2}}\\ + \frac{\rho'^2 \rho''}{\rho \sqrt{\varepsilon^2-\rho'^2}}-\frac{\rho''' \rho'}{(\varepsilon^2-\rho'^2)^{3/2}}-\frac{\rho''^2}{(\varepsilon^2-\rho'^2)^{3/2}}\\
-\frac{3\rho'^2 \rho''^2}{(\varepsilon^2-\rho'^2)^{3/2}}\Big)\Big)
\end{multline}
\begin{multline*}
\frac{d^2}{d\sigma^2}\frac{\partial \mathcal{L}}{\partial \rho''}=-\Big(\frac{\partial^2_{\rho}\varphi \rho'^2}{\sqrt{\varepsilon^2-\rho'^2}}+\frac{\partial_\rho \varphi \rho''}{\sqrt{\varepsilon^2-\rho'^2}}+ \frac{2\partial_\rho \varphi \rho'^2 \rho''}{(\varepsilon^2-\rho'^2)^{3/2}}
\\+ \frac{\rho''^2 \varphi}{(\varepsilon^2-\rho'^2)^{3/2}}+\frac{\varphi \rho' \rho''}{(\varepsilon^2-\rho'^2)^{3/2}}+\frac{3\varphi \rho'^2 \rho''^2}{(\varepsilon^2-\rho'^2)^{5/2}}\Big)
\end{multline*}
When $\varphi=Const.$ the action is a total derivative. So in the equation of motion only the non-constant part of $\varphi$ contributes. Now with $\varphi= C\rho^2 +...$, it is easy to see that $\rho'=0$ is not the solution to the Euler Lagrange equation. Moreover, one can check if we put $\rho'=\alpha$, where $\alpha$ is a non-zero constant, then the Euler Lagrange equation reduces to
\begin{equation}\label{G6}
C\sqrt{\varepsilon^2-\rho'^2}-\frac{C\rho'^2}{\sqrt{\varepsilon^2-\rho'^2}}=0
\end{equation}
which implies $\rho'=\frac{\varepsilon}{\sqrt{2}}$ is a solution to the equation of motion. But, notice that this solution is not periodic in $\sigma$. The equations are complicated and it is not possible to get a periodic solution by inspection.
\section{Dilaton as an external coupling}\label{H}
What we have done so far has subtle differences from the corresponding calculations in JT gravity for near-extremal black holes as in \cite{maldacena}. In those calculations, first of all, the integral over the dilaton in the bulk path integral can be done exactly. In the boundary Schwarzian action, the dilaton is therefore an external coupling that can be an arbitrary function of the boundary coordinate. Different choices of this function change the saddle points. For the dilaton being a constant, the boundary path integral can be done exactly.

In our computation, since we are working with spherically symmetric reduced gravity, the potential $V(\varphi)$ is much more complicated, and we are forced to do a one loop computation in the bulk. This also means that the boundary value of the dilaton in the boundary action is whatever it is for the classical solution, to leading order.

However, we now briefly discuss the Carlip boundary action with the dilaton being an external coupling
$\varphi(\sigma) = \varphi_{+} + A (\sigma)$. If the dilaton is a constant, the terms it multiplies in the action are a total derivative, so the contribution of the part $\varphi_{+}$ to the action just gives the Bekenstein Hawking entropy. The reason for writing the external coupling in this fashion is to explicitly display the entropy contribution.
\\
Let us consider the Carlip's action for the $DD$ case before splitting up the dilaton,
\begin{equation}\label{H1}
I_{bdry}^{DD}=\frac{1}{8\pi G}\int_{\Delta} d\sigma \varphi(\sigma)\Big(\tau'+\Big(\frac{\tau''}{\tau'^2}\Big)'\Big)
\end{equation}
In this whole section, for simplicity, we take $\beta$ equals to $2\pi$.
Using $\varphi = \varphi_+ + A(\sigma)$,
\begin{equation}\label{H2}
I_{bdry}^{DD}=\frac{\varphi_+}{4G} + \frac{1}{8\pi G} \int_{\Delta} d\sigma A(\sigma)\Big(\tau'+\Big(\frac{\tau''}{\tau'^2}\Big)'\Big)
\end{equation}
 It is easy to check that the equation of motion for the (\ref{H1}) action is
\begin{equation}\label{H3}
\varphi'+2\Big(\varphi' \frac{\tau''}{\tau'} \Big)' +\Big(\frac{\varphi'}{\tau'^2}\Big)''=0.
\end{equation}
We can now ask for what $\varphi(\sigma)$ is $\tau'=1$ the classical solution of equation of motion. This is true for $\varphi$ solving the equation
\begin{equation}\label{H4}
\varphi'''=-\varphi'
\end{equation}
Apart from $\varphi=\varphi_+$, the other solutions are $\sin \sigma$ and $\cos \sigma$. In spherically symmetric reduced gravity, the dilaton is positive (since it is like the squared radius of the sphere). We do not have a nice interpretation for $\varphi < 0$. But in principle, we can simplify the action for $\varphi = \sin \sigma$ (or cosine). However, we do not straightforwardly have Gaussian integrals, and we have coupling between adjacent Fourier modes. We will not have much more to say about this case.

For more general $\varphi(\sigma)$, let us consider the action for $DD$ boundary conditions.
\begin{equation}\label{H5}
I_{bdry}^{DD}=\frac{1}{8\pi G}\int_{\Delta} d\sigma \frac{\varphi(\sigma)}{\tau'}\Big(\tau'^2+\{\tau,\sigma\}-\frac{1}{2}\frac{\tau''^2}{\tau'^2}\Big)
\end{equation}
The action looks simpler written in terms of $\sigma(\tau)$ as the dynamical variable.
We use $\frac{d\tau(\sigma)}{d\sigma}=\frac{1}{\frac{d \sigma (\tau)}{d\tau}} $ and $\{\tau,\sigma\}=-\tau'^2 \{\sigma,\tau \}$.

In terms of $\sigma(\tau)$, the action is
\begin{equation}\label{H6}
I_{bdry}^{DD}=\frac{1}{8\pi G}\int_{\Delta} d\tau \varphi(\sigma)\Big(1-\{\sigma,\tau \}-\frac{1}{2}\sigma''^2\Big)
\end{equation}
In the above equation prime is a derivative with respect to $\tau$ . We can write the above expression in the following form :
\begin{equation}\label{H7}
I_{bdry}^{DD}=-\frac{1}{8\pi G}\int_{\Delta} d\tau \varphi(\sigma) \{\sigma,\tau \}
+\frac{1}{8\pi G}\int_{\Delta} d\tau \varphi(\sigma) \Big(1-\frac{1}{2}\frac{\sigma''^2}{\sigma'^2}\Big) .
\end{equation}
The first term is like the Schwarzian with external coupling $\varphi(\sigma)$. It is not easy to quantize either (\ref{H7}) for $\sigma(\tau)$ or (\ref{H5}) for $\tau(\sigma)$ in the presence of an arbitrary external coupling that is not a constant. However, one possible way is to first compute $Z_{ND}$, the partition function with fixed trace of extrinsic curvature and induced metric, which is discussed for JT gravity in (\cite{ALJZ}). Then the partition function we are after, $Z_{DD}$ can be obtained as the inverse Laplace transform of $Z_{ND}$.

One can similarly ask what will happen to Carlip's $DN$ action (\ref{a8}) when we switch the dynamical variable from $\tau(\sigma)$ to $\sigma(\tau)$. Using $\{\tau,\sigma\}=-\tau'^2 \{\sigma,\tau \}$  one can easily obtain
\begin{equation}\label{H8}
I^{DN}_{bdry}=\frac{1}{8 \pi G}\int_{\Delta} d\tau \varphi_{+} -\frac{\varepsilon}{16 \pi G}\int_{\Delta} d\tau \sigma' \partial_\rho \varphi\Big(1+ \{\sigma,\tau\}\Big)
\end{equation}
where $\varphi=\varphi_+ + \rho \partial_{\rho} \varphi$ and prime denotes a derivative with respect of the $\tau$ variable. In the $DN$ case, $\partial_\rho \varphi$ is fixed on the boundary, and therefore one of the choices is to make it constant. Now, with this choice, if we quantize this action to one loop with either the $\tau(\sigma)$ variable or the $\sigma(\tau)$ variable, we find that the one loop correction is divergent due to Gaussian-like integrals with the wrong sign, exactly as what we saw in previous sections. Upon doing an analytic continuation as in sections (\ref{s5}) and (\ref{s6}), the partition function has no $\beta$ dependence. Thus, there is no contribution from these boundary modes to the thermodynamics. Notice that when $\partial_{\rho}\varphi$ is constant, the value $\varphi$ will depend on the particular boundary curve via  $\varphi=\varphi_+ + \rho \partial_{\rho} \varphi$. One can also fix $\partial_{\rho} \varphi= \varepsilon B(\sigma)$ , where  $B(\sigma)$ is some function sourcing the Schwarzian action.
\section{Summary and discussion}
In this paper, we have studied the quantization of an interesting model of Carlip which provides a way of ascribing boundary degrees of freedom to a non-extremal black hole. This model has strong parallels to JT gravity which has been successfully used to associate boundary degrees of freedom to near-extremal black holes. Carlip's model is in the context of two dimensional dilaton gravity (spherically symmetric dimensionally reduced gravity). It introduces a stretched horizon which is allowed to fluctuate with the length fixed. In this paper, we have discussed the quantization of this model. The stretched horizon fluctuations, which reduce to time reparametrizations, are the boundary degrees of freedom in this model, exactly as in JT gravity. The bulk action contains the appropriate boundary terms required to make the variation of the action well-defined. The new feature in the quantization is that the path integral now contains an integral over time reparametrizations of the boundary in addition to the integral over the bulk fields. This is in the spirit of similar computations in JT gravity. Doing the bulk integrals in a one-loop approximation about the classical solutions, we see that the leading contribution to the boundary path integral comes from an integral over the boundary action with the boundary values of the classical metric and dilaton as input. We attempt to quantize this boundary action. Recognizing that the time reparametrization is an element of $Diff(S^1)/S^1 $, we use the symplectic form on this symplectic manifold to get a measure for the boundary path integral. This symplectic form comes from viewing $Diff(S^1)/S^1 $ as the coadjoint orbit of a coadjoint vector under the action of the Virasoro group. The measure depends on two constants, the coadjoint vector and the central charge. We find the classical solution to the boundary action in two cases: one with Dirichlet boundary conditions on the metric and the dilaton in the bulk path integral, and the other with the dilaton and its normal derivative fixed. We then consider one-loop fluctuations about the classical solution in both cases. We find that the fluctuations are Gaussians with the wrong sign and the one loop path integral is not finite. Either the classical solution is unstable or the action is indefinite, reminiscent of the conformal mode problem in quantum gravity. If the classical solution is unstable, this construction to ascribe boundary degrees of freedom to the non-extremal black hole has a problem. To cover the other possibility, we can perform an analytic continuation of the fluctuation to imaginary values (as was done by Gibbons, Hawking and Perry to deal with the conformal mode problem). We perform the analytic continuation and after appropriate zeta function regularizations, we find the one-loop correction to the path integral evaluated with the classical action. This correction is independent of the inverse temperature $\beta$, unlike in JT gravity. This is true whatever be the numerical value of the coadjoint vector or the central charge. So the correction to the path integral due to these boundary degrees of freedom does not contribute to the thermodynamics at least at one-loop. This is upon taking the integration measure to be that arising from the symplectic form. The bulk path integral, which is now non-trivial gives the entire contribution to the thermodynamics.
We contrast this to JT gravity where the thermodynamics of near-extremal black holes is entirely due to the boundary degrees of freedom in the path integral.

We also consider relaxing one of the assumptions in the model we studied. However, we then have a problem even finding classical solutions where $\rho(\sigma)$ is periodic in $\sigma$. Finally, we also discuss viewing the dilaton as an external coupling, as was done in JT gravity. Again, for the case where we can quantize the action with the dilaton as external coupling, we find that the one-loop correction to the classical contribution in the path integral is not finite, and we need to analytically continue the field modes to imaginary values. Upon then quantizing, the path integral is again independent of $\beta$ and does not contribute to thermodynamics.

\section{Acknowledgements} We would like to thank Steve Carlip for a very useful discussion and for his detailed comments on a draft of this paper. MA also acknowledges the Council of Scientific and Industrial Research (CSIR), Government of India for financial assistance.
\section{Appendix A}\label{A}

We want to analyze equation (\ref{a3a}) for the static case $i.e (\partial_{\tau} \phi= 0 )$ with metric $g_{ab}= e^{2\xi(\rho,\tau)}(d \rho^2 + \rho^2 d\tau^2)$. The equation (\ref{a3a}) for $\rho \tau$ will become\\
\begin{equation}\label{A1}
\partial_{\rho}\partial_{\tau}\varphi-\partial_{\tau}\xi \partial_{\rho}\varphi-\Big(\frac{1}{\rho} + \partial_{\rho}\xi\Big)\partial_{\tau}\varphi=0 .
\end{equation}
Since $\varphi$ is static, $\partial_{\tau}\xi \partial_{\rho}\varphi$ must vanish. This implies that either $\varphi$ should be just a constant, which is the trivial case or $\partial_\tau \xi= 0$. We will consider the latter.
The equation (\ref{a3a}) for $\rho \rho$ and $\tau \tau$ is
\begin{eqnarray}\label{A2}
&& \nabla_{\tau}\nabla_{\tau}\varphi= \frac{1}{2}g_{\tau \tau} V(\varphi);\nonumber \\
&& \nabla_{\rho}\nabla_{\rho}\varphi= \frac{1}{2}g_{\rho \rho} V(\varphi).
\end{eqnarray}
Using the fact that $g_{\tau \tau}= \rho^2 g_{\rho \rho}$ and the above equations, we get
\begin{equation}\label{A3}
\rho^2 \nabla_{\rho}\nabla_{\rho}\varphi =\nabla_{\tau}\nabla_{\tau}\varphi .
\end{equation}
Upon simplification, it becomes
\begin{equation}\label{A4}
\partial^2_{\rho}\varphi -2 \partial_{\rho}\xi \partial_{\rho}\varphi -\frac{1}{\rho}\partial_{\rho}\varphi=0
\end{equation}
On solving the above equation one can easily obtain
\begin{equation}\label{A5}
\partial_{\rho}\varphi= \tilde C \rho e^{2\xi(\rho)}
\end{equation}
where $\tilde C$ is a constant. Thus the leading $\rho$ dependence of $\varphi$ beyond the constant term for all $\xi$ which are well behaved and which can be written as power series in $\rho$ about $\rho = 0$ is quadratic. Note that the equation (\ref{A5}) must be obeyed with any allowed potential $V(\varphi)$. Of course, $\varphi$ must also solve (\ref{A2}). But this tells us that for a wide class of potentials $V(\varphi)$, the dilaton $\varphi = \varphi_{+} + C \rho^2 +.....$ to leading order in $\rho$.
\section{Appendix B}\label{B}
The general solution to a general 2-D dilaton model  is \cite{Cavaglia}
\begin{equation}\label{B1}
ds^2=-\Big(N(\varphi)-M\Big)dt^2+\frac{1}{N(\varphi)-M}d\phi^2
\end{equation}
where,
\begin{equation}\label{B2}
N(\varphi)=\int_{0}^{\varphi'}V(\varphi')d\varphi'
\end{equation}
For spherical symmetric dimensionally reduced solutions of general relativity
\begin{equation}\label{B3}
V(\varphi)=\frac{1}{2\sqrt{\varphi}}
\end{equation}
For the spherically symmetric solution the metric takes following form,
\begin{equation}\label{B4}
ds^2=-\Big(\sqrt{\varphi}-M\Big)dt^2+\frac{1}{\sqrt{\varphi}-M}d\phi^2
\end{equation}
After doing Wick rotation and setting $\varphi=\frac{r^2}{4}$ and $r_+=2M$, we get
\begin{equation}\label{B5}
ds_{E}^2=\frac{1}{2} \Big( (r-r_+)d\tau^2+\frac{r^2}{r-r_+}dr^2 \Big)
\end{equation}
Now let us make a coordinate transformation $r=r_+ +\frac{\kappa \rho^2}{2}$, where $\kappa=\frac{1}{2r_+}$,
\begin{equation}\label{B6}
ds_{E}^2=\frac{1}{4\kappa}\Big(\kappa \rho^2 d\tau^2 +(1+\kappa^4 \rho^4 +2\kappa^2 \rho^2)d\rho^2\Big)
\end{equation}
We can break the above metric as $g^{(0)}_{ab}$ which is the flat metric and $g^{(1)}_{ab}$ as correction to the metric. Under the assumption that $\rho$ is small, we can expand the connection $\Gamma= \Gamma^{(0)} +\Gamma^{(1)}$, where $\Gamma^{(0)}$ is with respect to $g^{0}$ and $\Gamma^{(1)}$ is the leading correction. It can be checked that
\begin{align}\label{B7}
\Gamma^{\tau (0)}_{\rho \tau}&=\frac{1}{\rho} & \Gamma^{\rho (0)}_{\tau \tau}&=-\kappa \rho\sim O(\varepsilon)\\
\Gamma^{\rho (1)}_{\rho\rho} & \sim O(\varepsilon) & \Gamma^{\rho (1)}_{\tau \tau}&\sim O(\varepsilon^3)
\end{align}
 Other connection symbols are zero. Let $\sigma$ be the coordinate on the stretched horizon. The induced metric $\varepsilon^2$ on the stretched horizon is,
 \begin{equation}\label{B8}
 \varepsilon^2=\kappa \rho^2 \tau'^2+ X\rho'^2
 \end{equation}
 where $X=1+2\kappa^2 \rho^2+ \kappa^4 \rho^4$. The tangent and normal vector to the stretched horizon is,
 \begin{align}\label{B9}
 t^a&=\frac{1}{\varepsilon}\Big(\rho',\tau'\Big) & n^a=\frac{\rho \kappa}{\varepsilon \sqrt{X}}\Big(\tau',-\frac{\rho'X}{\kappa^2 \rho^2}\Big)
 \end{align}
 We can expand $n^a= n^a_{(0)}+n^a_{(1)}$ where $n^a_{(0)}\sim O(1)$ and $n^a_{(1)}\sim O(\epsilon^2)$. The extrinsic curvature of the stretched horizon is
 \begin{equation}\label{B10}
 K=g_{bc}t^a t^b \nabla_a n^c=g_{bc}t^a t^b \nabla_a n^c_{(0)}+g_{bc}t^a t^b \nabla_a n^c_{(1)}
 \end{equation}
 Since $K$ contains terms which are either proportional to $n^a$ or $t^b\partial_b n^a$, $n_{(1)}$ will only contribute in higher order terms. We only want terms of order $\varepsilon$. So we will only concentrate on the $n_{(0)}$ term.
 \begin{equation}\label{B11}
 K=(g^{(0)}_{bc}+g^{(1)}_{bc})t^a t^b (\partial_a n^c_{(0)}+(\Gamma^{c(0)}_{ad}+\Gamma^{c(1)}_{ad})n^d_{(0)}).
 \end{equation}
 Therefore,
 \begin{equation}\label{B12}
 K=K^{(0)} +g^{(0)}_{bc}t^a t^b\Gamma^{c(1)}_{ad}n^d_{(0)}+g^{(1)}_{bc}t^a t^b\Gamma^{c(1)}_{ad}n^d_{(0)}+(\partial_a n^c_{(0)}+\Gamma^{c(0)}_{ad}n^d_{(0)})g^{(1)}_{bc}t^a t^b
 \end{equation}
 where $K^{(0)}\sim O(\varepsilon)$ is Carlip's extrinsic curvature. Before going further notice that
 \begin{align}\label{B13}
 \rho & \sim O(\varepsilon), & \rho'&\sim O(\varepsilon^2), & \tau'&\sim O(1), & \tau''&\sim O(\varepsilon)
 \end{align}
 \begin{align*}
 t^\rho &\sim O(\varepsilon) & t^\tau &\sim O(1/\varepsilon)  & n_{(0)}^{\rho} &\sim O(1)  &  n_{(0)}^{\tau}&\sim O(1)\\
 g^{(0)}_{\rho \rho}& \sim O(1)   & g^{(0)}_{\tau \tau}& \sim O(\varepsilon^2) & g^{(1)}_{\rho \rho}&\sim O(\varepsilon^2)
 \end{align*}
 In the above expression, the first line in the equations are Carlip's assumptions. Now to compute order of correction to the Carlip's extrinsic curvature, consider (\ref{B12}) term by term,
 \begin{equation}\label{B14}
 g^{(0)}_{bc}t^a t^b\Gamma^{c(1)}_{ad}n^d_{(0)}=g^{(0)}_{\rho \rho}t^\rho t^\rho\Gamma^{\rho(1)}_{\rho \rho}n^\rho_{(0)}+g^{(0)}_{\rho \rho}t^\rho t^\tau\Gamma^{\rho(1)}_{\tau \tau}n^\tau_{(0)}.
 \end{equation}
 Using (\ref{B13}) it is easy to see that each term in the above expression is $O(\varepsilon^3)$. Therefore the above term contributes at $O(\varepsilon^3)$. To see what would have been the correction due to $n_{(1)}$, we can replace $n_{(0)}$ with $n_{(1)}$ which will increase the order by $\varepsilon^2$. Therefore, the $n_{(1)}$ correction contributes at $O(\varepsilon^5)$. Similarly
 \begin{equation}\label{B15}
 g^{(1)}_{bc}t^a t^b\Gamma^{c(1)}_{ad}n^d_{(0)}=g^{(1)}_{\rho \rho}t^\rho t^\rho\Gamma^{\rho(1)}_{\rho \rho}n^\rho_{(0)}+g^{(1)}_{\rho \rho}t^\rho t^\tau \Gamma^{\rho(1)}_{\tau \tau}n^\tau_{(0)}.
 \end{equation}
 Each term in the above expression is $O(\varepsilon^5)$. To see what would have been the correction due to $n_{(1)}$, replace $n_{(0)}$ with $n_{(1)}$ which will increase the order by $\varepsilon^2$. Therefore the $n_{(1)}$ correction contribute at $O(\varepsilon^7)$. Now let us analyze the last term in (\ref{B12}) --- for that let us first consider
 \begin{equation}\label{B16}
 \Gamma^{c(0)}_{ad}n^d_{(0)}g^{(1)}_{bc}t^a t^b=  \Gamma^{\rho(0)}_{\tau \tau}n^\tau_{(0)}g^{(1)}_{\rho \rho}t^\tau t^\rho .
 \end{equation}
 Using (\ref{B13}), the above term contributes at $O(\varepsilon^5)$. To see what would have been the correction due to $n_{(1)}$, replace $n_{(0)}$ with $n_{(1)}$ which will increase the order by $\varepsilon^2$. Therefore $n_{(1)}$ correction contribute at $O(\varepsilon^7)$. Now we are just left with
 \begin{equation}\label{B17}
 g^{(1)}_{bc}t^a t^b\partial_a n^c_{(0)}=g^{(1)}_{\rho \rho}t^\rho t^b \frac{\partial_{\sigma n^{\rho}_{(0)}}}{\varepsilon}
 \end{equation}
 It is easy to check the above term is of $O(\varepsilon^3)$. To see what would have been the correction due to $n_{(1)}$, replace $n_{(0)}$ with $n_{(1)}$ which will increase the order by $\varepsilon^2$. Therefore the $n_{(1)}$ correction contributes at $O(\varepsilon^5)$. Therefore the extrinsic curvature is
 \begin{equation}\label{B18}
 K= K^{(0)}+ O(\varepsilon^3)
 \end{equation}
\section{Appendix C} \label{C}
In this appendix, our task is to find a function $H_{\alpha}$ such that $\iota_{V_\alpha} \omega = dH_{\alpha}$.
We simplify the expression for $\omega$ in (\ref{b1}) to
\begin{eqnarray}
\omega = \int_{0}^{2\pi} d\sigma [- \frac{c}{48\pi} \frac{d\tau'}{\tau'} \wedge \partial_{\sigma}(\frac{d\tau'}{\tau'}) - \frac{4\pi^2}{\beta^2}b_0 d\tau\wedge \partial_{\sigma}d\tau .
\label{b3}
\end{eqnarray}

Let
\begin{eqnarray}
\omega_1 = \int_{0}^{2\pi} d\sigma - (\frac{c}{48\pi}) \frac{d\tau'}{\tau'} \wedge \partial_{\sigma}(\frac{d\tau'}{\tau'}) \nonumber \\
\omega_2 = \int_{0}^{2\pi} d\sigma \frac{4\pi^2}{\beta^2}(- b_0) d\tau\wedge \partial_{\sigma}d\tau .
\label{b4}
\end{eqnarray}
We use
\begin{eqnarray}
\iota_{V_{\alpha}} (\beta \wedge \gamma) = (\iota_{V_{\alpha}} \beta) \wedge \gamma + (-1)^p \beta \wedge (\iota_{V_{\alpha}} \gamma),
\label{b5}
\end{eqnarray}
for $p$ form $\beta$ and $q$ form $\gamma$.
Consider $\iota_{V_{\alpha}}\omega_1$. This is
\begin{eqnarray}
\iota_{V_{\alpha}}\omega_1 &=& \int_{0}^{2\pi} d\sigma (- \frac{c}{48\pi}) \frac{\partial_{\sigma} (\alpha \tau')}{\tau'}
\partial_{\sigma} (\frac{d\tau'}{\tau'})  \nonumber \\
&+& (-1) \int_{0}^{2\pi} d\sigma (- \frac{c}{48\pi}) (\frac{d\tau'}{\tau'}) \partial_{\sigma} (\frac{\partial_{\sigma} (\alpha \tau')}{\tau'}).
\label{b6}
\end{eqnarray}
We now integrate the second integral by parts (with boundary terms going to zero) to get
\begin{eqnarray}
\iota_{V_{\alpha}}\omega_1 &=& \int_{0}^{2\pi} d\sigma (- 2\frac{c}{48\pi}) [\alpha' + \alpha \frac{\tau''}{\tau'} ]
\partial_{\sigma} [\frac{d\tau'}{\tau'}] = dH_{\alpha 1}
\label{b7}
\end{eqnarray}
where
\begin{eqnarray}
H_{\alpha 1} = -\frac{c}{48\pi}\int_{0}^{2\pi} d\sigma~ [2\alpha' \frac{\tau''}{\tau'} + \alpha (\frac{\tau''}{\tau'})^2].
\label{b8}
\end{eqnarray}
Similarly,
\begin{eqnarray}
\iota_{V_{\alpha}}\omega_2 &=& -\int_{0}^{2\pi} d\sigma \frac{4\pi^2}{\beta^2} b_0 \alpha \tau' \partial_{\sigma}d\tau \nonumber \\
&+& \int_{0}^{2\pi} d\sigma b_0 d\tau \partial_{\sigma}(\alpha \tau').
\label{b9}
\end{eqnarray}
After integration by parts, this is
\begin{eqnarray}
\iota_{V_{\alpha}}\omega_2 &=& - 2\int_{0}^{2\pi} d\sigma \frac{4\pi^2}{\beta^2}b_0 \alpha \tau' d\tau' = dH_{\alpha2},
\label{b10}
\end{eqnarray}
where
\begin{equation}
H_{\alpha2} = - \int_{0}^{2\pi} d\sigma \frac{4\pi^2}{\beta^2}b_0 \alpha \tau'^2.
\label{b11}
\end{equation}
Finally, after several integrations by parts, we get
\begin{eqnarray}
H_{\alpha} = H_{\alpha1} + H_{\alpha2} = \int_{0}^{2\pi} d\sigma \alpha [ \frac{2c}{48\pi}\{\tau,\sigma\} - \frac{4\pi^2}{\beta^2}b_0 \tau'^2 ].
\label{b12a}
\end{eqnarray}
\section{Appendix D}\label{D}
\subsection{1- loop action about arbitrary classical solution}
Just for completeness we want to write 1-loop action about any arbitrary classical solution. Let $\tau= \tau_c$ is some classical solution, then the perturbation about the classical solution can be written as $\tau=\tau_c + \delta \tau$. Consider Carlip's action
\begin{equation}\label{D.1}
I^{non}_{bdry}=-\frac{1}{16\pi G}\int_{\Delta}d\sigma \frac{\varepsilon  }{\tau'^2}\partial_{\rho}\varphi\{\tau'^2 -\{\tau,\sigma\}\}
\end{equation}
since
\begin{equation}\label{D.2}
\frac{\{\tau,\sigma\}}{\tau'^2}=\frac{\tau'''}{\tau'^3}-\frac{3}{2}\frac{\tau''^2}{\tau'^4}
\end{equation}
It is straightforward to show
\begin{multline}\label{D.3}
\frac{\{\tau,\sigma\}}{\tau'^2}=\frac{\{\tau_c,\sigma\}}{\tau_c'^2}+\Big(\frac{\delta \tau'''}{\tau_c'^3}-\frac{3\delta \tau' \tau_c'''}{\tau_c'^4}-\frac{3\tau_c'' \delta \tau''}{\tau_c'^4} +\frac{\delta \tau' \tau_c''}{\tau_c'^4}\Big)\\+\Big(\frac{6\delta \tau'^2\tau_c'^3}{\tau_c'^5}-\frac{3 \delta \tau''^2}{2\tau_c'^4}+\frac{12 \delta \tau' \tau_c'' \delta \tau''}{\tau_c'^5}-\frac{15 \tau_c''^2 \delta \tau'^2}{\tau_c'^6}-\frac{3\delta \tau' \delta \tau'''}{\tau_c'^4}\Big)
\end{multline}
Now the action looks like
\begin{multline}\label{D.4}
I^{non}_{bdry}=-\frac{1}{16\pi G}\int_{\Delta}d\sigma \varepsilon \partial_{\rho}\varphi\Big[1-\frac{\{\tau_c,\sigma\}}{\tau_c'^2}-\Big(\frac{\delta \tau'''}{\tau_c'^3}-\frac{3\delta \tau' \tau_c'''}{\tau_c'^4}-\frac{3\tau_c'' \delta \tau''}{\tau_c'^4} +\frac{\delta \tau' \tau_c''}{\tau_c'^4}\Big)\\-\Big(\frac{6\delta \tau'^2\tau_c^3}{\tau_c'^5}-\frac{3 \delta \tau''^2}{2\tau_c'^4}+\frac{12 \delta \tau' \tau_c'' \delta \tau''}{\tau_c'^5}-\frac{15 \tau_c''^2 \delta \tau'^2}{\tau_c'^6}-\frac{3\delta \tau' \delta \tau'''}{\tau_c'^4}\Big)\Big] .
\end{multline}
If we take $\partial_{\rho}\varphi=C\rho^2$ and $\rho=\varepsilon/\tau'$ then
\begin{multline}\label{D.5}
I^{non}_{bdry}=-\frac{2C\varepsilon^2}{16\pi G}\int_{\Delta}d\sigma \Big\{\Big[\frac{1}{\tau_c'}-\frac{\{\tau_c,\sigma\}}{\tau_c'^3}\Big]+\Big[\frac{\delta \tau'}{\tau_c'^2}-\Big(\frac{\delta \tau'''}{\tau_c'^4}-\frac{4\delta \tau' \tau_c'''}{\tau_c'^5}-\frac{3\tau_c'' \delta \tau''}{\tau_c'^5} +\frac{15\delta \tau' \tau_c''}{\tau_c'^6}\Big)\Big]\\+\Big[\frac{\delta \tau'^2}{\tau_c'^3}-\Big(\frac{10\delta \tau'^2\tau_c'^3}{\tau_c'^6}-\frac{3 \delta \tau''^2}{2\tau_c'^5}+\frac{15 \delta \tau' \tau_c'' \delta \tau''}{\tau_c'^6}-\frac{45 \tau_c''^2 \delta \tau'^2}{2\tau_c'^7}-\frac{4\delta \tau' \delta \tau'''}{\tau_c'^5}\Big)\Big]\Big\}.
\end{multline}
The first term corresponds to the value of action at classical solution. The second term has to be zero once the equation of motion for the classical solution is imposed and the third term is the 1-loop action. One can trivially see that $\tau'= Const$ is a solution (the second term will vanish). Also notice that, the sign of the third term depends non-trivially on the details of the classical solution.

\section{Appendix E}\label{E}
\subsection{Zeta function regularization}
Using the zeta function one can regularize divergent series and integrals. The zeta function is defined as
\begin{equation}\label{E1}
\zeta(s)=\sum_{n=1}^{\infty}\frac{1}{n^s}
\end{equation}
for $Re(s)>1$ and is given by its analytic continuation elsewhere. Values of $\zeta{s}$ for various $s$ are known, for example, $\zeta(0)$ and $\frac{d\zeta}{ds}|_{s=0}$ are
 \begin{equation}\label{E2}
 \zeta(0)= 1+1+1+...=-\frac{1}{2}
 \end{equation}
 and
 \begin{equation}\label{E3}
 \zeta'(0)=-\sum_{n=1}^{\infty} \log n = -\frac{1}{2} \log{2\pi}
 \end{equation}
Now consider the following infinite product
\begin{equation}\label{E4}
\prod_{n=1}^{\infty} (\alpha n^x).
\end{equation}We can write it as
\begin{equation}\label{E5}
\prod_{n=1}^{\infty} (\alpha n^x)= \exp{\Big(\sum_{n=1}^{\infty} \log (\alpha n^x) \Big)}.
\end{equation}
Using property  $\log (ab)= \log a + \log b$ and (\ref{E5}),  we get
\begin{equation}\label{E6}
\prod_{n=1}^{\infty} (\alpha n^x)= \exp{\Big(\log \alpha \sum_{n=1}^{\infty} 1 + x \sum_{n=1}^{\infty} \log n \Big)}.
\end{equation}
Now using (\ref{E2}),(\ref{E3}) and (\ref{E6}), we obtain
\begin{equation}\label{E7}
\prod_{n=1}^{\infty} (\alpha n^x)= \exp{\Big(-\frac{1}{2}\log \alpha  + \frac{x}{2}\log{2\pi} \Big)}.
\end{equation}
Therefore, using zeta function summation
\begin{equation}\label{E8}
\prod_{n=1}^{\infty} (\alpha n^x)= \sqrt{\frac{(2\pi)^{x}}{\alpha}}
\end{equation}
\subsection{Calculation of $\#_{DN}$, $\#_{DD}$  and $Pf(\omega)|_{\beta=2\pi}$}
We can write
\begin{equation}\label{E11}
\prod_{n=1}^{\infty} \Big(\frac{n^4}{b^2}-n^2\Big)=\exp{\Big \{\sum_{n=1}^{\infty}\Big(\log{\frac{1}{b^2}}+2\log{n}+\log{(n^2-b^2)}\Big)\Big\}}.
\end{equation}
In the expression above, the use of  $\log AB= \log A + \log B $ inside the infinite sum can give rise to multiplicative anomaly in the infinite product while using zeta function regularization \cite{Elizalde}. But we will show that anomaly vanishes in the cases we are interested in this subsection. It has already been shown in \cite{Elizalde} that there is no multiplicative anomaly in the infinite product expression (\ref{E8}) in the previous subsection. Using $\zeta(0)$ and $\zeta'(0)$, we get
\begin{equation}\label{E12}
\prod_{n=1}^{\infty} \Big(\frac{n^4}{b^2}-n^2\Big)= 2\pi b\exp{\Big \{\sum_{n=1}^{\infty}\Big(\log{(n^2-b^2)}\Big)\Big\}}
\end{equation}
Using the identity
\begin{equation}\label{E13}
\sum_{n=1}^{\infty}\frac{1}{n^2-a^2}=\frac{1-a\pi \cot{(a\pi)}}{2a^2},
\end{equation}
multiplying both the sides with $2a$ and integrating with respect to $a$ from 0 to $b$ yields
\begin{equation}\label{E14}
\sum_{n=1}^{\infty}\log{(n^2-b^2)}=\log{\Big(\frac{2\sin{b\pi}}{b}\Big)} \mbox{\hspace{2mm} when} \sin{b\pi}\geq 0 .
\end{equation}
Using (\ref{E12}) and (\ref{E14}), we get
\begin{equation}\label{E15}
\prod_{n=1}^{\infty} \Big(\frac{n^4}{b^2}-n^2\Big)=4\pi\sin{b\pi}\mbox{\hspace{2mm} when} \sin{b\pi}\geq 0 .
\end{equation}
In equation (\ref{21}), $b^2=2/5$ for which $\sin{\frac{\sqrt{2}\pi}{\sqrt{5}}}>0$. Therefore
\begin{equation}\label{E16}
\prod_{n=1}^{\infty} \Big(\frac{5 n^4}{2}-n^2\Big)=4\pi\sin{\frac{\sqrt{2}\pi}{\sqrt{5}}}
\end{equation}
\begin{equation}\label{E17}
\#_{DN}=\prod_{n=1}^{\infty} \Big(\frac{5 n^4}{2}-n^2\Big)^{-1}=\frac{1}{4\pi\sin{\frac{\sqrt{2}\pi}{\sqrt{5}}}}
\end{equation}
One can derive an identity similar to (\ref{E14}),
\begin{equation}\label{E18}
\sum_{n=1}^{\infty}\log{(n^2+b^2)}=\log{\Big(\frac{2\sinh{b\pi}}{b}\Big)}
\end{equation}
Since
\begin{equation}\label{E18A}
\#_{DD}= \prod_{n=1}\frac{1}{n^2(n^2+\frac{1}{2})}
\end{equation}
which can be written as
\begin{equation}\label{E18B}
\#_{DD}=\exp{\Big(-\sum_{n=1} \log{n^2}-\sum_{n=1}\log{\Big(n^2+\frac{1}{2}\Big)}\Big)},
\end{equation}
using the identity in (\ref{E14}) and (\ref{E8}), we get
\begin{equation}\label{E18C}
\#_{DD}= \frac{1}{4\sqrt{2}\pi} \frac{1}{\sinh{\frac{\pi}{\sqrt{2}}}}.
\end{equation}
Now, we can easily derive $\mu$ using the above identities, since Pfaffian $Pf(\omega)|_{\beta=2\pi}$ is
\begin{equation}\label{E19}
Pf(\omega)|_{\beta=2\pi}=\prod_{n=1}^{\infty}\Big(\frac{ic}{24}\Big) \Big(n^3+\frac{48\pi b_0 n}{c}\Big)
\end{equation}
and
\begin{equation}\label{E19A}
\mu= Pf(\omega)|_{\beta=2\pi}\prod_{n=1}(2i)
\end{equation}
Therefore
\begin{equation}\label{E19B}
\mu=\prod_{n=1}^{\infty}\Big(\frac{-c}{12}\Big) \Big(n^3+\frac{48\pi b_0 n}{c}\Big)
\end{equation}
which can be written as
\begin{equation}\label{E20}
\mu=\exp{\Big(\sum_{n=1} \log{\Big(\frac{-cn}{12}\Big)}+\sum_{n=1} \log{\Big(n^2+\frac{48\pi b_0 }{c}\Big)}\Big)}.
\end{equation}
Using identity (\ref{E18}) and (\ref{E8})
\begin{equation}\label{E21}
\mu = i\sqrt{\frac{2}{b_0}}\sinh{\Big(\pi \sqrt{\frac{48\pi b_0}{c}}\Big)}
\end{equation}
In the above expression $b_0>0$. When $b_0<0$ the above expression will become
\begin{equation}\label{E22}
\mu =i \sqrt{\frac{2}{|b_0|}}\sin{\Big(\pi \sqrt{\frac{48\pi |b_0|}{c}}\Big)}
\end{equation}
\subsubsection{Multiplicative anomaly}\label{E.1}
Zeta function regularization should be used with care as pointed out before. The use of  $\log AB= \log A + \log B $ inside an infinite sum can give rise to a multiplicative anomaly in infinite products while using zeta regularization. It is nicely discussed in \cite{Elizalde}. Here we want to show that for the cases done in the above subsection, this anomaly vanishes. Let us say we want to compute
\begin{equation}\label{E.11}
\prod_{n=1}^{\infty} \lambda_n \mu_n= \exp\Big({\sum_{n=1}^{\infty}}\log(\lambda_n \mu_n)\Big)
\end{equation}
now,
\begin{equation}\label{E.12}
{\sum_{n=1}^{\infty}}\log(\lambda_n \mu_n)={\sum_{n=1}^{\infty}}\log(\lambda_n)+{\sum_{n=1}^{\infty}}\log(\mu_n) +\alpha
\end{equation}
where $\alpha$ is the multiplicative anomaly. In our case, $\lambda_n=(n^2 + b^2)$ and $\mu_n = n^2$.
\begin{equation}\label{E.13}
{\sum_{n=1}^{\infty}}\log(n^2(n^2 + b^2))={\sum_{n=1}^{\infty}}\log(n^2 + b^2)+{\sum_{n=1}^{\infty}}\log(n^2) +\alpha(b) .
\end{equation}
This is because the anomaly coefficient $\alpha$ can depend in general on $b$. Now when $b=0$, $\lambda_n=n^2 $ and $\mu_n = n^2$ and therefore there will be no anomaly, as has been shown in \cite{Elizalde}. This implies $\alpha(0)=0$. Using the results from the last subsection\footnote{${\sum_{n=1}^{\infty}}\log(n^2)=\log 2\pi$ and ${\sum_{n=1}^{\infty}}\log(n^2 + b^2)= \log\Big(\frac{2 \sinh(\pi b)}{b}\Big)$}, we can write (\ref{E.13}) as
\begin{equation}\label{E.14}
{\sum_{n=1}^{\infty}}\log(n^2(n^2 + b^2))= \log\Big(\frac{4\pi \sinh(\pi b)}{b}\Big) +\alpha(b)
\end{equation}
Now, define $F(b)= {\sum_{n=1}^{\infty}}\log(n^2(n^2 + b^2))$, then
\begin{equation}\label{E.15}
\frac{dF}{db}=2b {\sum_{n=1}^{\infty}}\frac{n^2}{(n^2(n^2 + b^2))}
\end{equation}
Using ${\sum_{n=1}^{\infty}}\frac{n^2}{(n^2(n^2 + b^2))}= \frac{1}{2b}\Big(\pi\coth(\pi b)-\frac{1}{b}\Big)$ in the above equation\footnote{It can be easily computed using contour integral techniques.}, we get
\begin{equation}\label{E.16}
\frac{dF}{db}= \Big(\pi\coth(\pi b)-\frac{1}{b}\Big)
\end{equation}
also notice that,
\begin{equation}\label{E.17}
\frac{d}{db} \log\Big(\frac{4\pi \sinh(\pi b)}{b}\Big)= \Big(\pi\coth(\pi b)-\frac{1}{b}\Big)
\end{equation}
This implies $\frac{d\alpha(b)}{db}=0$. Now together with the fact that $\alpha(0)=0$, it implies $\alpha(b)=0$. Hence the anomaly vanishes in this case. By taking $b$ to $ib$ in this sub-section, we can also derive that the anomaly vanishes for other infinite products we encounter in the DN case. Using the result that when $\lambda_n= c \mu_n^\delta$ where $c,\delta \in R$, there will be no multiplicative anomaly \cite{Elizalde} in the other infinite products in the expression for the partition function, such as that in the previous subsection.

\end{document}